\documentclass[a4paper,fleqn,usenatbib]{mnras}
\usepackage{newtxtext,newtxmath}

\usepackage{ae,aecompl}
\usepackage{graphicx}	
\usepackage{amsmath}	
\usepackage{mathtools}

\usepackage{hyperref}
\hypersetup{draft=false}

\title[DEVILS: $\sigma_{\mathrm{SFR}}$-M$_{\star}$ evolution]{Deep Extragalactic VIsible Legacy Survey (DEVILS): Evolution of the $\sigma_{\mathrm{SFR}}$-M$_{\star}$ relation and implications for self-regulated star formation}
\author[L. J. M. Davies et. al.]{L. J. M. Davies$^{1}$\thanks{E-mail:
 luke.j.davies@uwa.edu.au}, J. E. Thorne$^{1}$, S. Bellstedt$^{1}$,  M. Bravo$^{1}$, A. S. G. Robotham$^{1}$, S. P. Driver$^{1}$, \newauthor R. H. W. Cook$^{1}$, L. Cortese$^{1,2}$, J. D'Silva$^{1}$, M. W. Grootes$^{3}$, B.W. Holwerda$^{4}$, A. M. Hopkins$^{5}$, \newauthor M. J. Jarvis$^{6,7}$, C. Lidman$^{8}$, S. Phillipps$^{9}$, M. Siudek$^{10,11}$ \\
$^{1}$ICRAR, The University of Western Australia, 35 Stirling Highway, Crawley, WA 6009, Australia\\
$^{2}$ ARC Centre of Excellence for All Sky Astrophysics in 3 Dimensions (ASTRO 3D), Australia\\
$^{3}$Netherlands eScience Center, 1098 XG Amsterdam, The Netherland\\
$^{4}$Physics \& Astronomy Department, University of Louisville, 40292 KY, Louisville, USA\\
$^{5}$Australian Astronomical Optics, Macquarie University, 105 Delhi Road, North Ryde, NSW 2113, Australia\\
$^{6}$Astrophysics, University of Oxford, Denys Wilkinson Building, Keble Road, Oxford OX1 3RH, UK\\
$^{7}$Department of Physics and Astronomy, University of the Western Cape, Robert Sobukwe Road, Bellville 7535, South Africa\\
$^{8}$The Research School of Astronomy and Astrophysics, Australian National University, ACT 2601, Australia\\
$^{9}$Astrophysics Group, School of Physics, University of Bristol, Tyndall Avenue, Bristol BS8 1TL, UK\\
$^{10}$Institut de F\'{\i}sica d'Altes Energies (IFAE), The Barcelona Institute of Science and Technology, 08193 Bellaterra (Barcelona), Spain\\
$^{11}$National Centre for Nuclear Research, ul. Pasteura 7, 02-093, Warsaw, Poland
}

\date{Accepted XXX. Received YYY; in original form ZZZ}

\pubyear{2016}

\begin{document}
\label{firstpage}
\pagerange{\pageref{firstpage}--\pageref{lastpage}}
\maketitle

\begin{abstract}
We present the evolution of the star-formation dispersion - stellar mass relation ($\sigma_{\mathrm{SFR}}$-M$_{\star}$) in the DEVILS D10 region using new measurements derived using the \textsc{ProSpect} spectral energy distribution fitting code. We find that $\sigma_{\mathrm{SFR}}$-M$_{\star}$  shows the characteristic `U-shape' at intermediate stellar masses from $0.1<z<0.7$ for a number of metrics, including using the deconvolved intrinsic dispersion. A physical interpretation of this relation is the combination of stochastic star-formation and stellar feedback causing large scatter at low stellar masses and AGN feedback causing asymmetric scatter at high stellar masses. As such, the shape of this distribution and its evolution encodes detailed information about the astrophysical processes affecting star-formation, feedback and the lifecycle of galaxies. We find that the stellar mass that the minimum $\sigma_{\mathrm{SFR}}$ occurs evolves linearly with redshift, moving to higher stellar masses with increasing lookback time and traces the turnover in the star-forming sequence. This minimum $\sigma_{\mathrm{SFR}}$ point is also found to occur at a fixed specific star-formation rate (sSFR) at all epochs (sSFR$\sim10^{-9.6}$\,yr$^{-1}$). The physical interpretation of this is that there exists a maximum sSFR at which galaxies can internally self-regulate on the tight sequence of star-formation. At higher sSFRs, stochastic stellar processes begin to cause galaxies to be pushed both above and below the star-forming sequence leading to increased SFR dispersion. As the Universe evolves, a higher fraction of galaxies will drop below this sSFR threshold, causing the dispersion of the low-stellar mass end of the star-forming sequence to decrease with time.

\end{abstract}


\section{Introduction}

A ubiquitous feature of the star-forming galaxy population is the tight correlation between the rate at which they are forming new stars (star formation rate, SFR) and their total baryonic mass currently in stars (stellar mass, M$_{*}$). The SFR-M$_{\star}$ relation, commonly referred to as the star-forming `main sequence'  \citep[SFS,][]{Elbaz07,Noeske07,Salim07,Whitaker12, Johnston15, Davies16b} is found to exist for galaxies over a range of epoch and environments, and has been shown by numerous studies to remain roughly linear out to the early Universe, but with increasing normalisation as a function of lookback time \citep[$e.g.$][]{Schreiber15, Lee15, Thorne21}. The observational linearity and low scatter along this relation has been been interpreted as evidence that the majority of star-forming galaxies exist in a self-regulated equilibrium state \citep[$i.e.$][]{Bouche10, Daddi10, Genzel10, Lagos11, Lilly13, Dave13, Mitchell16}. In this model, the inflow rate of gas for future star-formation is balanced by the rate at which new stars are formed and the outflow of gas from feedback events ($i.e.$ Supernovae, SNe, and Active Galactic Nuclei, AGN). 

However, within the full distribution of galaxies the relationship between star-formation and stellar mass in not this simplistic. Galaxies that sit off the tight locus of the SFS are unlikely to fit within this simple self-regulated model. For example,  populations such as star-bursting sources which lie above the SFS, the passive cloud which sits below the SFS and `green valley' sources which sit between the SFS and passive cloud, are all likely undergoing evolution through the SFR-M$_{\star}$ plane, which does not follow this self-regulated model. Moreover, various studies have found that even within the SFS galaxies are constantly changing position due to small star-burst/quenching events \citep[$i.e.$][]{Magdis12, Tacchella16}. This means that while the locus of the SFS remains constant at a given epoch and evolves smoothly with time, individual galaxies are constantly moving above and below the SFS locus, resulting in the observed intrinsic scatter. The SFS also shows a number of characteristics that suggest this simple self-regulated model make break down in specific regimes. Many studies have measured the non-linearity and flattening of the SFS at high stellar masses \citep[$e.g.$][]{Rodighiero10, Elbaz11, Whitaker12,Lee15,Katsianis16, Grootes17, Grootes18, Thorne21}, and various astrophysical processes have been invoked that drive the most massive galaxies away from the linear SFS \citep[$e.g.$][]{Erfanianfar16,Abramson14,Willett15,Cook19}. As such, understanding the astrophysical origin of galaxies across the full SFR-M$_{\star}$ plane can provide key insights in the evolutionary processes that are diving galaxy properties. 

To fist order, the position of a galaxy within this plane is governed by its star-formation history \citep[SFH,][]{Madau98, Kauffmann03, Bellstedt20}, $i.e.$ the rate at which stars were formed as a function of time, and the availability of gas for star-formation episodes. In turn these properties can be fundamentally altered by the primary events that occur in a galaxies life, such as SNe- \citep{Dekel86, DallaVecchia08, Scannapieco08} and AGN-feedback \citep{Kauffmann04,Fabian12}, environmental processes such as strangulation, stripping and starvation \citep[$e.g.$][]{Giovanelli85, Moore99, Peng10, Cortese11, Darvish16, Cortese21}, morphological evolution \citep{Conselice14, Eales15}, gas accretion events adding new fuel \citep{Kauffmann06,Sancisi08, Mitchell16}, and mergers \citep[$e.g.$][and see review of Conselice 2014]{Bundy04, Baugh06, Kartaltepe07, Bundy09,Jogee09,deRavel09,Lotz11,Robotham14}. Combined it is these events that shape a galaxy's SFHs and ultimately result in the distribution within the SFR-M$_{\star}$ plane \citep[][]{Abramson16, Caplar19}.

One of the key diagnostics in the variation of SFHs that leads to the distribution of galaxy properties is the dispersion along the SFR-M$_{\star}$ relation \citep[$\sigma_{\mathrm{SFR}}$-M$_{\star}$,][]{Guo15, Willett15,Katsianis19, Davies19a}. At a given stellar mass, this dispersion essentially encodes the variation in galaxy SFHs - caused by the processes outlined above. Recently numerous studies have explored the shape of the $\sigma_{\mathrm{SFR}}$-M$_{\star}$ relation, finding varied observational measurements of the dispersion at a given stellar mass and redshift \citep{Elbaz07, Noeske07, Rodighiero10,Whitaker12, Schreiber15, Guo15, Santini17}. However, in the local Universe a consensus picture is arising that the $\sigma_{\mathrm{SFR}}$-M$_{\star}$ relation appears to be `U-shaped', with high dispersion at both low and high stellar masses and a minimum dispersion point at around log$_{10}$[M/M$^{*}]$=9-10 \citep{Willett15, Davies19a}.  In \cite{Davies19a} we explored the variation in the measured $\sigma_{\mathrm{SFR}}$-M$_{\star}$ relation for different star-forming population selection techniques and SFR indicators using the Galaxy And Mass Assembly \citep[GAMA,][]{Driver11,Hopkins13,Liske15,Driver16, Baldry18} sample. We found that this `U-shaped' $\sigma_{\mathrm{SFR}}$-M$_{\star}$ relation is ubiquitous irrespective of selection method and SFR indicator, suggesting that this shape is fundamental to the galaxy population and encodes information about the astrophysical processes that are driving the position of galaxies in the  SFR-M$_{\star}$ plane.    

Simulations can also offer further insights into the $\sigma_{\mathrm{SFR}}$-M$_{\star}$ relation and the physical processes driving its evolution. \cite{Sparre15} use the Illustris simulation \citep{Vogelsberger14} to explore the SFS and at $z\sim0$ find a relatively flat $\sigma_{\mathrm{SFR}}$-M$_{\star}$ at 9$<$log$_{10}$[M$_{*}$/M$_{\odot}$]$<$10.5 and increasing dispersion to higher masses, roughly consistent with observations. \cite{Katsianis19} applied a similar approach to EAGLE \citep{Crain15, Schaye15, McAlpine16, Matthee18} and find a `U-shaped' $\sigma_{\mathrm{SFR}}$-M$_{\star}$ relation similar to that of \cite{Willett15}. Conversely, \cite{Matthee18} undertake a similar study with EAGLE but find a linearly decreasing $\sigma_{\mathrm{SFR}}$-M$_{\star}$ with stellar mass. However, it must be noted that both of these studies apply a SFR cut to their samples, and the overall measurement of $\sigma_{\mathrm{SFR}}$-M$_{\star}$ is very sensitive to the exact choice of SFR cut used \citep{Davies19a}  

Following the work of \cite{Davies19a}, we now expand of analysis the $\sigma_{\mathrm{SFR}}$-M$_{\star}$ relation to explore its evolution out to $z\sim0.8$ using the Deep Extragalactic VIsible Legacy Survey \citep[DEVILS,][]{Davies18}. Importantly DEVILS is designed as an intermediate $0.3<z<1$ counterpart to GAMA, using the same selection, measurement and analysis techniques, allowing us to draw direct comparisons between galaxy samples over the last $\sim$8\,Gyr of universal history. After parametrising the evolution of the $\sigma_{\mathrm{SFR}}$-M$_{\star}$ relation, we then use its changing shape, and the simulation predictions regarding the astrophysical nature of the various features of this relation, to suggest a lower stellar mass and/or specific SFR limit at which self-regulated, main-sequence star-formation begins to give way to stochastic stellar-feedback/star-formation at the low stellar mass end.    \\

\section{Data}

\begin{figure}
\begin{center}
\includegraphics[scale=0.6]{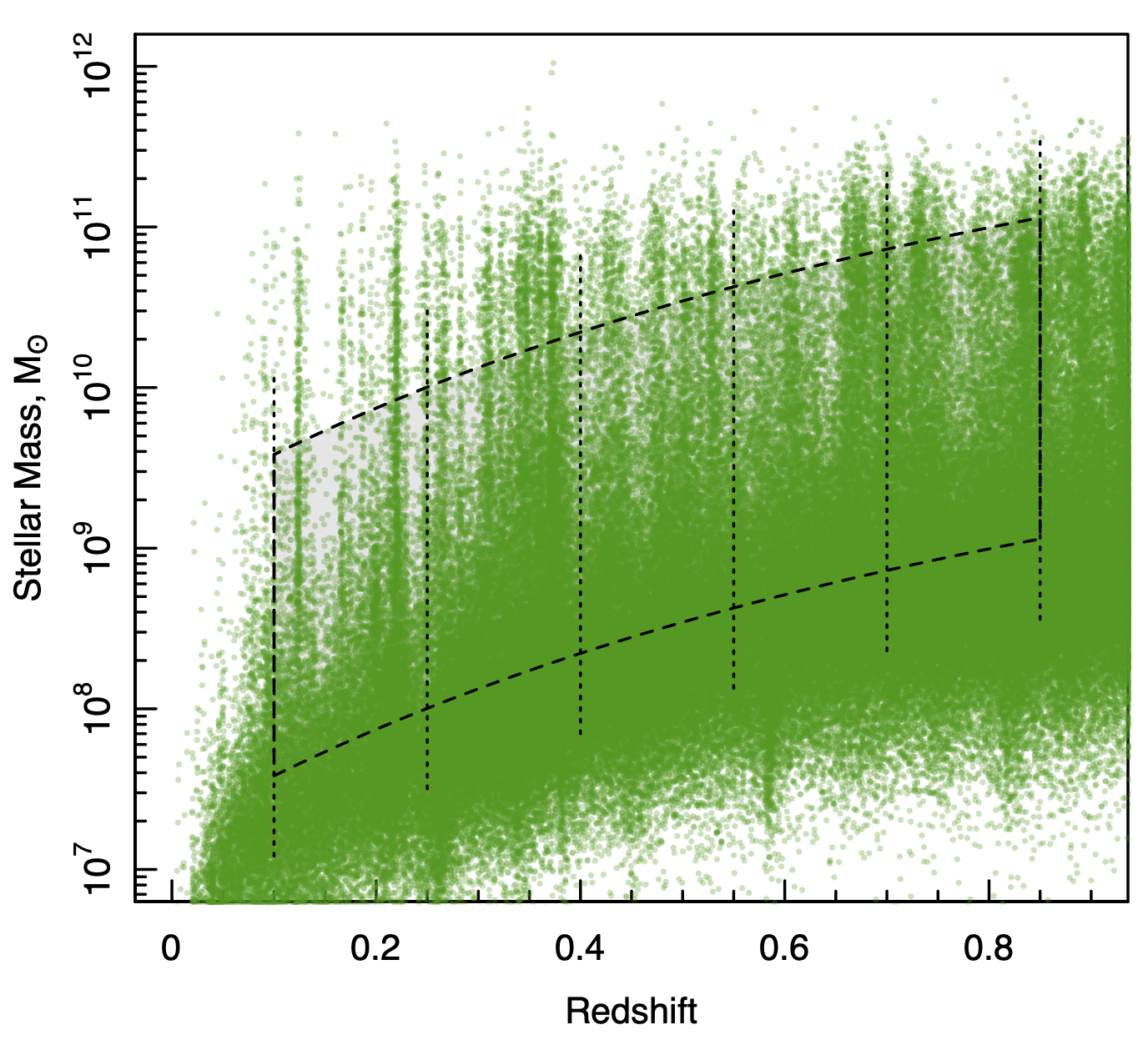}

\caption{The distribution of points in the M$_{\star}$-$z$ plane from DEVILS galaxies using the sample outlined in \citet{Thorne21} with boxes over plotted to define the sample selection used in this work. The dotted vertical lines show the separation between redshift bins used in our analysis. We split the sample into five $\Delta z=0.15$ bins between 0.1<z<0.85. The lower dashed line displays the $g-i$ colour completeness limit presented in \citet{Thorne21} for a rest-frame colour-complete sample as a function of redshift, while the upper dashed line is this relation plus 2\,dex (to exclude to most massive galaxies which will be under-sampled in this relatively small area field). These line bound the region used in our polynomial fitting (see Figure \ref{fig:Msig}).}
\label{fig:Mz}
\end{center}
\end{figure}

\begin{figure*}
\begin{center}
\includegraphics[scale=0.425]{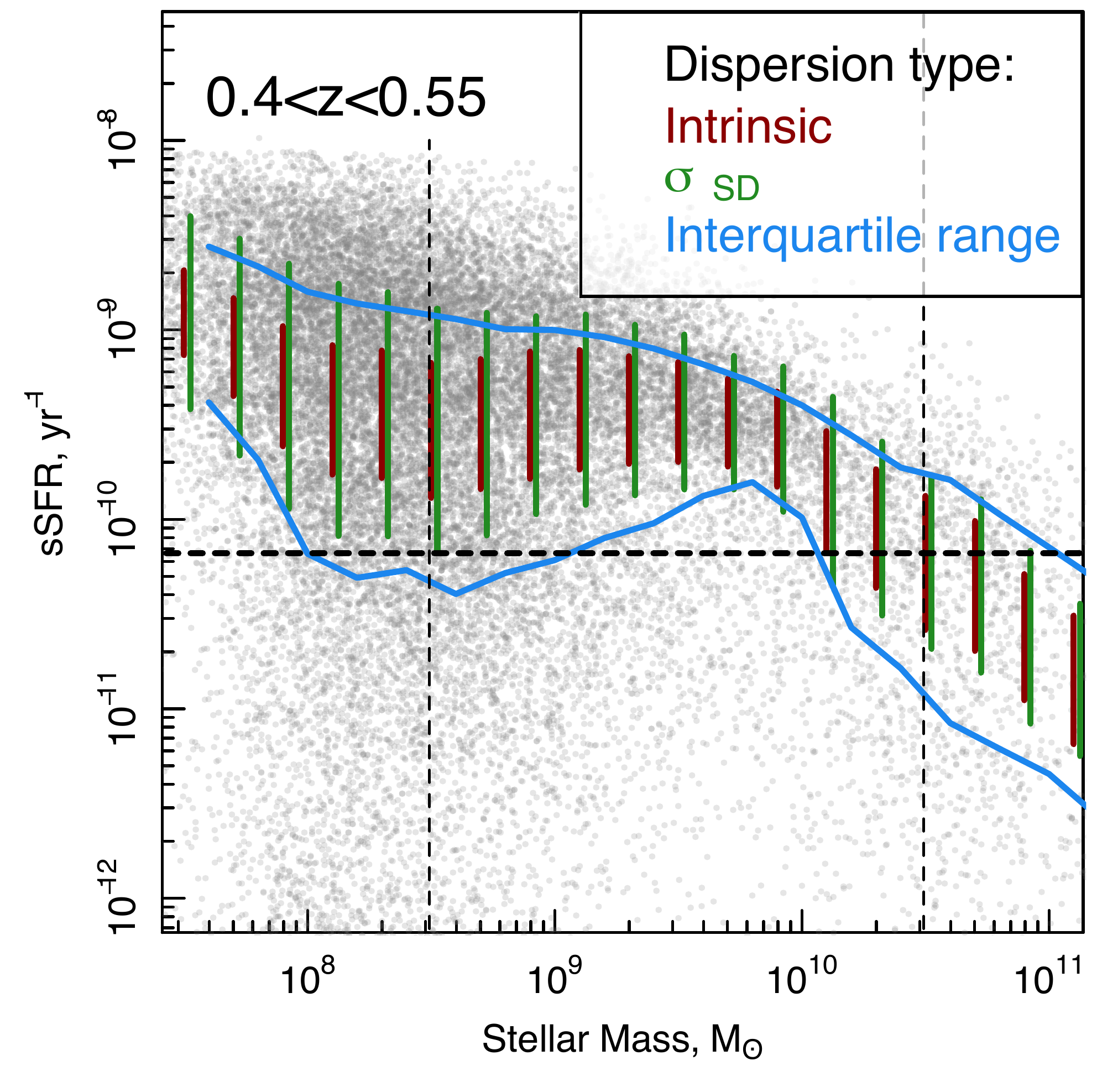}
\includegraphics[scale=0.68]{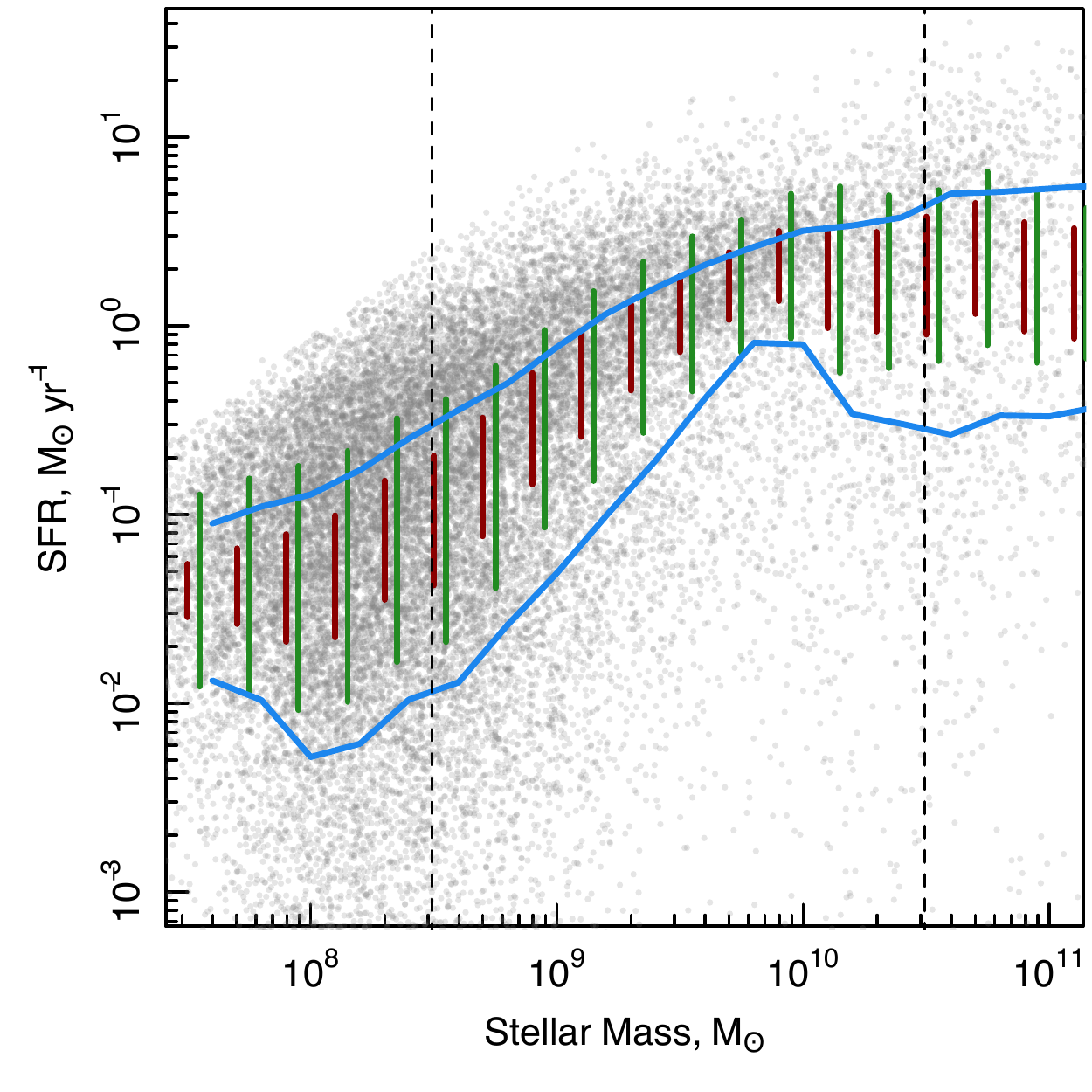}

\caption{Example of dispersion measurement along the sSFR-M$_{\star}$ (left) and SFR-M$_{\star}$ (right) relations in a single redshift bin at $0.4<z<0.55$ using the D10-\textsc{ProSpect} sample.  We split the distribution into log$_{10}$[M$_{\star}$]=0.2\,dex bins and measure both the standard deviation ($\sigma_{SD}$, green lines), interquartile range (bounded by blue lines), and intrinsic scatter (red lines) at each stellar mass. We highlight that while the distributions appear skewed to have more points above the $\sigma_{SD}$,  interquartile range and intrinsic scatter values than below, this is due to the fact that many passive systems fall at much lower SFRs/sSFRs than shown here.  The `U-shaped' distribution is visible in all dispersion metrics, with minimum at $M^{*}\sim10^{10}$M$_{\odot}$. The fitting range displayed in Figure \ref{fig:Mz} is shown as the dashed vertical lines. The dotted horizontal line in the left panel displays the sSFR cut that lies at 1\,dex below the main sequence normalisation point measured at 9$<$log$_{10}$[M*/M$_{\odot}$]$<$10. This is used in Section \ref{sec:sigma_M_cut}. Note that while there is a cutout to the maximum sSFR derived in the \textsc{ProSpect} analysis, this occurs below our stellar mass limit.}
\label{fig:M-SFR}
\end{center}
\end{figure*}

\subsection{The Deep Extragalactic VIsible Legacy Survey}

Briefly, DEVILS is an ongoing spectroscopic survey being undertaken with the Anglo-Australian Telescope (AAT). The survey aims to build a high completeness ($>$85\%) sample of $\sim$60,000 galaxies to Y$<$21.2\,mag in three well-studied deep extragalactic fields: D10 (COSMOS), D02 (ECDFS) and D03 (XMM-LSS). This sample will provide the first high completeness sample at intermediate redshift, allowing for the robust parametrisation of group and pair environments in the distant Universe. The science goals of the project are varied, from the environmental impact on galaxy evolution at intermediate redshift, to the evolution of the halo mass function over the last $\sim$7\,billion years. For full details of the survey science goals, survey design, target selection, photometry and spectroscopic observations see \cite{Davies18, Davies21a}.

The DEVILS regions were specifically chosen to cover areas with extensive existing and oncoming imaging campaigns to facilitate broad range of science topics. In this work we only use the DEVILS D10 region which represents a sub-region of the Cosmic Evolution Survey region \citep[COSMOS][]{Scoville07}, covering 1.5deg$^{2}$ of the UltraVISTA \cite{McCracken12} field and centred at R.A.=150.04, Dec=2.22. This field is prioritised for DEVILS early science as it is the most spectroscopically complete, has the most extensive multi-wavelength coverage of the DEVILS fields, and has already been processed to derive robust galaxy properties through spectral energy distribution (SED) fitting, see below.

\begin{figure*}
\begin{center}
\includegraphics[scale=0.6]{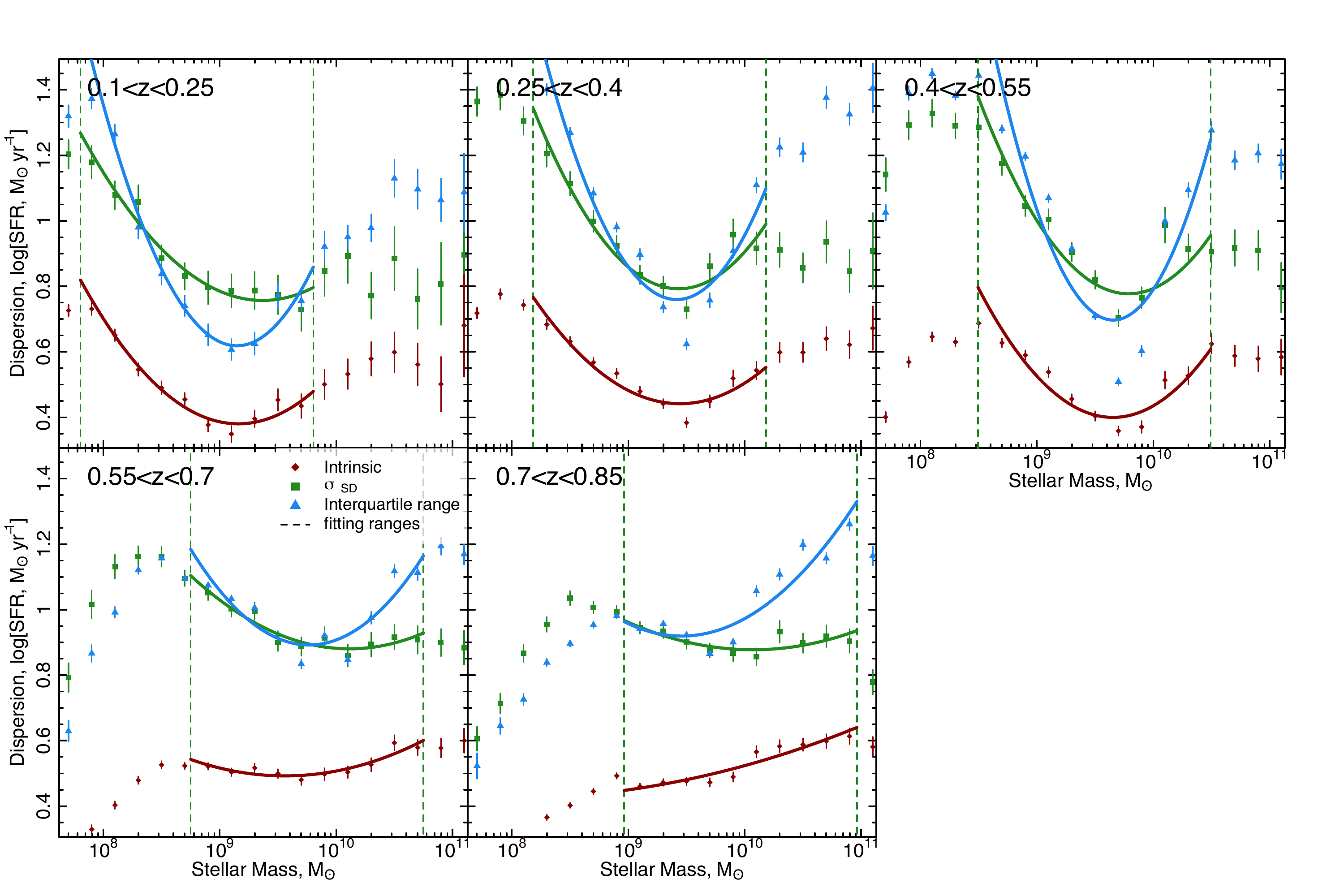}

\caption{The dispersion along the SFR-M$_{\star}$ relation in different redshift bins using both standard deviation ($\sigma_{SD}$, green) and interquartile range (blue). Errors are calculated from 100 bootstrap resamples of the distribution. Dashed vertical lines are the sample ranges described in Figure \ref{fig:Mz}. We fit the distributions using a least squares 2$^{nd}$ order polynomial regression between the sample limit lines (solid lines). Note $\sigma_{SD}$ and interquartile range agree at the low stellar mass end but differ at above M$_{\star}$$\gtrsim10^{10}$M$_{\odot}$. This is indicative of the scatter being a $log-normal$ at low stellar masses (from stochastic SF and stellar feedback) and asymmetric at high stellar masses (from AGN-feedback), see \citep{Davies19a}. All distributions show the characteristic `U-shaped' distribution between the sample limits. }
\label{fig:Msig}
\end{center}
\end{figure*}

\subsection{Sample selection and fitting ranges}

In this work we utilise the derived galaxy properties and redshifts for galaxies in the D10 region discussed in \cite{Thorne21}. Briefly, \cite{Thorne21} fit galaxies in the D10 region using the \textsc{ProSpect} \citep{Robotham20} SED fitting code to derive galaxy properties such as stellar mass, SFR, star-formation history and metallicity. They discuss the validity of these measurements and then perform a detailed analysis of the evolution of the SFR-M$^*$ relation and stellar mass function. As such, we do not go into any further detail here, but refer the reader to \cite{Thorne21} for a detailed description of the values used in this work. We note that for this paper we use the DEVILS-internal D10-\textsc{ProSpect} catalogue \texttt{DEVILS\_D10ProSpectCat\_02\_02\_2021\_v0.3}.

In order to explore the evolution of the $\sigma_{\mathrm{SFR}}$-M$_{\star}$ relation in DEVILS we first split our sample into five $\Delta z=0.15$ redshift bins between $0.1<z<0.85$. Below this redshift we use the results from \cite{Davies19a} from GAMA (which probes a much larger, and therefore representative volume), while above this redshift, it becomes somewhat difficult to parameterise the shape of the $\sigma_{\mathrm{SFR}}$-M$_{\star}$ relation as the \cite{Thorne21} sample becomes significantly incomplete to low stellar mass galaxies. These redshift bins are shown as the dashed vertical lines in Figure \ref{fig:Mz}.    

Following this we only consider the $\sigma_{\mathrm{SFR}}$-M$_{\star}$ relation for a stellar mass complete sample of galaxies at each redshift bin. To determine this range, \cite{Thorne21} calculate the rest-frame $g-i$ colour completeness limit, $M_{\mathrm{lim}}$, as a function of stellar mass and lookback time as:

\begin{equation}
\mathrm{log}_{10}(M_{\mathrm{lim}}/M_{\odot})=\frac{1}{4}t_{\mathrm{lb}}+7.25
\end{equation}

\noindent where $t_{\mathrm{lb}}$ is lookback time in Gyrs. This line essentially represents the lower stellar mass at which the sample is complete to both red-passive and blue-starfoming galaxies at a given epoch. Figure \ref{fig:Mz} displays this as the lower dashed line. At the upper stellar mass end, the D10 region will also be incomplete to the most massive systems due to the small volume probed in the local Universe. To conservatively overcome this we limit our sample to systems that have log$_{10}(M^{\mathrm{*}}/M_{\odot})<$log$_{10}(M_{\mathrm{lim}}+2.0)$. While the 2\,dex range here is somewhat arbitrary, we show in Section \ref{sec:Mmin} that this selection bounds the minimum SFR dispersion point in the $\sigma_{\mathrm{SFR}}$-M$_{\star}$ relation. In addition, we find that removing this upper cutoff does not significantly change any of the results presented in this paper.  In subsequent figures we will display the full sample in each redshift bin ($i.e.$ without these stellar mass completeness cuts imposed) but show the ranges of the stellar mass complete sample as dashed vertical lines, and only fit our data between these lines.         

\begin{figure*}
\begin{center}
\includegraphics[scale=0.8]{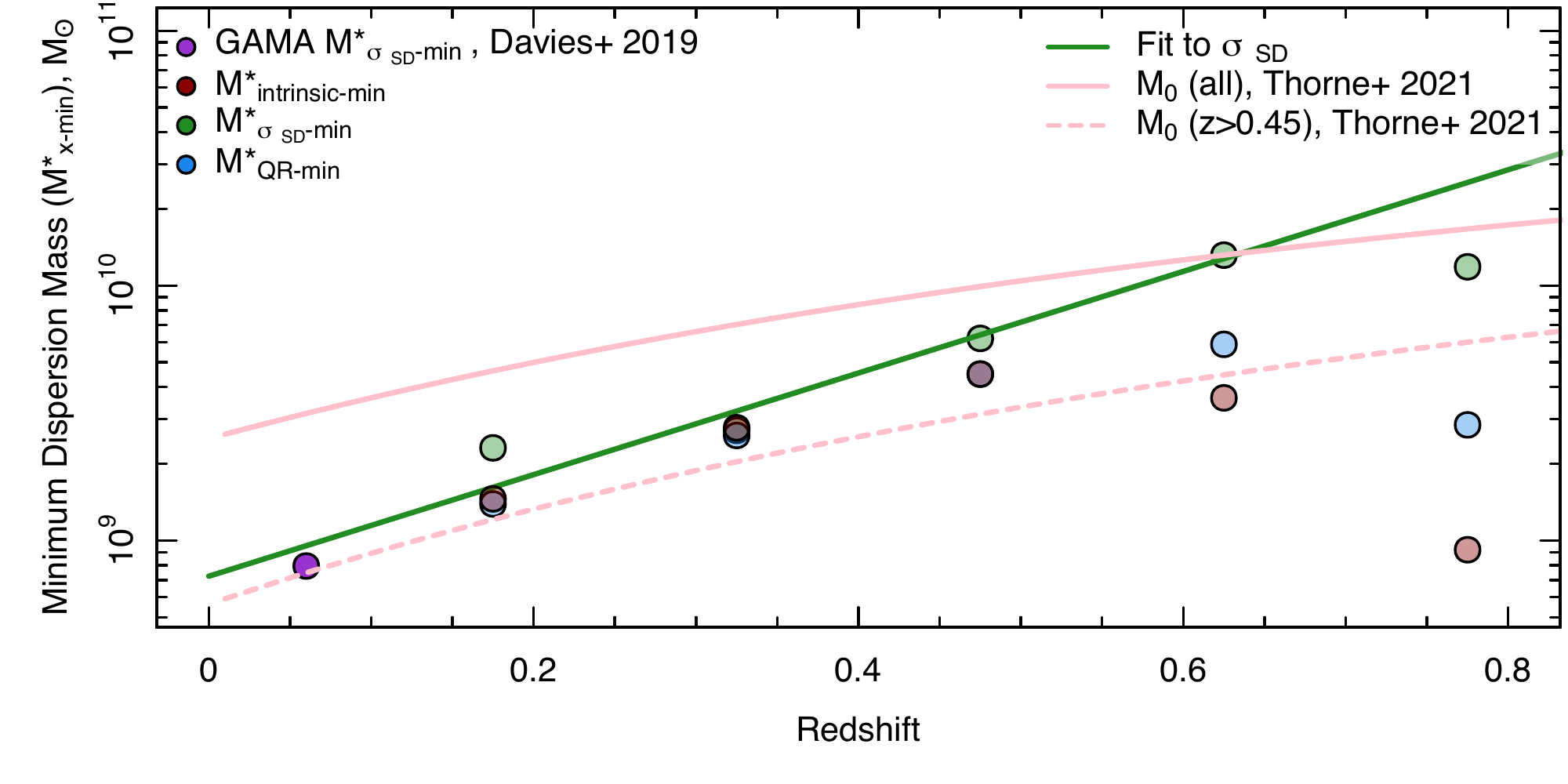}

\caption{Evolution of the minimum dispersion stellar mass point (M$^{*}_{x-min}$) with redshift. M$^{*}_{x-min}$ is presented for our polynomial fits for $\sigma_{SD}$, interquartile range and the intrinsic scatter. We also over-plot the minimum point for the GAMA sample at z$\sim0.05$ from \citet{Davies19a}. For all methods the minimum dispersion point increases in stellar mass with redshift. We fit the GAMA data point and $\sigma_{SD}$-fit values to produce the solid green line, which parametrises the evolution of the minimum dispersion point with redshift. We also show the M$_{0}$ evolution for both the full sample and just $z>0.45$ galaxies from \citet{Thorne21}, respectively. M$_{0}$ represents the `break point' where the star-forming main-sequence flattens at the high mass end. We find that the evolution of M$_{0}$ is similar to the evolution of M$^{*}_{x-min}$, potentially suggesting a common origin. }
\label{fig:redevol}
\end{center}
\end{figure*}

Finally, within our analysis we must also decide whether to parametrise the dispersion in our sample in terms of SFR or sSFR. This is somewhat a matter of personal preference, as both quantities are of interest and both can be used to define the dispersion about the star-forming main sequence. More broadly, both properties have merits in terms of exploring the galaxy population and its evolution, and previous studies exploring the scatter along the main-sequence have almost equally split between the choice of SFR or sSFR. While SFR is a more direct measure of the current activity within the galaxy and the conversion rate of gas into stars, the sSFR is more indirectly a measure of the relative growth rate of galaxies, and the energy input into the system per unit mass. Largely this choice here is down to the specific question being asked. Given that we wish to not only parameterise the measured scatter on the population but also intrinsic scatter (and associated errors), we decide to primarily use the metric with the smallest measurement error. As sSFR contains co-variant errors from both SFR and stellar mass (which must be combined in quadrature), we will use dispersion measurements for SFR. However, we note that in our analysis we do measure the dispersion and reproduce all figures for sSFRs as well. We find that the results are almost identical in both cases, and therefore this choice does not affect our conclusions in any way. Therefore, we only opt to show our dispersion metrics and their evolution in terms of SFR for the rest of this paper.

\section{The $\sigma_{\mathrm{SFR}}$-M$_{\star}$ relation}
\label{sec:sigma}

We next determine the $\sigma_{\mathrm{SFR}}$-M$_{\star}$ relation in each of our redshift bins. We note here that in this initial analysis we do not perform any sub-selection for star-forming/passive systems, but undertake our analysis for the full galaxy sample. We split the sample at each redshift into 23 $\Delta$log$_{10}$(M$_{\star}$/M$_{\odot}$)=0.2 bins between 7.2$<$log$_{10}$(M$_{\star}$/M$_{\odot}$)$<$12.0. Then following \cite{Davies19a}, we measure the dispersion of SFRs in each stellar mass bin using three different metrics. 

First, we calculate the standard deviation, here $\sigma_{SD}$, in log$_{10}$(SFR) in each bin. Second, we calculate the interquartile range in each bin ($i.e.$ making no assumption about the shape of the distribution). Finally, we also calculate the intrinsic scatter in each bin using the \textsc{hyperfit} \citep{Robotham15} package with full MCMC optimisation, assuming a unimodal, one dimensional distribution of SFRs. This is achieved by, in each stellar mass and redshift bin, assigning random x-values to sources in each bin and fitting for SFR with errors. Errors on each SFR are taken from \citet{Thorne21}, and for reference the median SFR error as a function of stellar mass and SFR for our $0.4<z<0.55$ sample are shown in Figure \ref{fig:errors}.  

This measurement aims to remove the component of the dispersion that is driven by measurement error - and is therefore most likely to represent the true astrophysical properties of the $\sigma_{\mathrm{SFR}}$-M$_{\star}$ relation. For our measurement errors per source, we take the upper and lower bound of the SED-derived properties outlined in \cite{Thorne21} as a 1$\sigma$ error for both SFR and stellar mass.      

Figure \ref{fig:M-SFR} displays an example of our three dispersion metrics for a single epoch at $0.4<z<0.55$ . We over-plot the $\sigma_{SD}$, interquartile range and intrinsic scatter measurements in each mass bin. The stellar mass completeness ranges discussed in the previous section are displayed as the dashed vertical lines. From this Figure alone, it is clear that the $\sigma_{\mathrm{SFR}}$-M$_{\star}$ relation at this epoch displays the characteristic `U-shape' when considering all dispersion metrics, with large dispersion at the low stellar mass end, a low dispersion pinch-point at intermediate stellar masses and a return to larger dispersion at high stellar masses.

In Figure \ref{fig:Msig} we then display the $\sigma_{\mathrm{SFR}}$-M$_{\star}$ relation for $\sigma_{SD}$, interquartile range and intrinsic scatter dispersion metrics as a function of stellar mass in each of our five redshift bins. We note that  each panel covers a significant redshift range and therefore some of the observed dispersion could be due to evolution in SFRs across this range. However, we also repeat our analysis splitting each redshift range in two and find that our results do not change (but errors are increased).  

Errors on $\sigma_{SD}$ and the interquartile range are calculated in the same way as \cite{Davies19a}. Briefly, for both metrics we perform a 100 bootstrap resamplings within the upper and lower bounds of the \textsc{ProSpect} SED fitting ranges in both stellar mass and SFR, and re-calculate the dispersion in each resampling. The error then represents the standard deviation of the dispersions in each resample. This is intended to take into account the varying measurement error in each of our SFR indicators as a function of stellar mass. For our $\sigma_{SD}$ error, we then also include a statistical error calculated as:

\begin{equation}
\label{eq:errors}
\mathrm{Err}_{\sigma_{\mathrm{SFR}_i}}\sim\frac{\sqrt{2\sigma_{\mathrm{SFR}_i}^4 (N_i-1)^{-1}}}{2\sigma_{\mathrm{SFR}_i}}
\end{equation}

\noindent where $i$ is the index of the stellar mass bin and $N$ is the number of galaxies in that bin \citep{Rao73}. We then combine this in quadrature with the error calculated from our bootstrap resamples. For errors on the intrinsic scatter, these are directly obtained in \textsc{hyperfit} from the MCMC posterior chains, where we input the 1$\sigma$ errors for both SFR and stellar mass.
                 
We find that, particularly between our stellar mass completeness lines (dashed verticals in Figure \ref{fig:Msig}), the dispersion shows the characteristic `U-shape' for all dispersion metrics, but that this begins to break down in our highest redshift bin, where either our samples become too incomplete to parameterise the shape or the `U-shape' of this distribution no longer applies (see discussion in Section \ref{sec:discuss}). Interestingly, we also find that the $\sigma_{SD}$ metic and interquartile range metric are largely consistent in dispersion measurement in the `U-shape' region and at the low mass end, but diverge for high stellar masses. This is consistent the proposed origin of the dispersion along the SFS, in which at the low stellar mass end, stochastic star-formation and stellar feedback is likely to induce symmetric $log-normal$ scatter about the SFS, as galaxies would be both enhanced in SFR through star-burst events, and suppressed in SFR through stellar feedback events. In the $log-normal$ regime $\sigma_{SD}$ and the interquartile range should be similar (interquartile range $\sim1.35\sigma_{SD}$), as we observe. However, at the high stellar mass end, processes that drive galaxies off the SFS (AGN feedback, etc) are likely predominantly quenching in nature, and once galaxies fall off the sequence they very rarely return ($i.e.$ this is a one-directional non-stochastic process). As such, this leads to a distribution of SFRs that is not $log-normal$ in distribution (potentially $log-normal$ with a power-law tail or bimodal). In this regime $\sigma_{SD}$ and interquartile range will not provide the same dispersion measurement, as seen in Figure \ref{fig:Msig} \citep[this is discussed in more detail in][as similar results are seen in GAMA, and further in Section \ref{sec:discuss}]{Davies19b}. As expected, the intrinsic scatter shows a much lower dispersion than the other metrics, but still shows the same `U-shape' within our sample limits at $z<0.7$. We then fit the dispersion values between our sample ranges using a simple 2$^{nd}$-order polynomial using a least-squares regression to parameterise the the parabolic `U-shape'. These are displayed as the coloured solid lines in Figure \ref{fig:Msig}.

\subsection{M$^{*}_{x-min}$ - the minimum dispersion point and its evolution} 
\label{sec:Mmin}

To parameterise the evolution of the $\sigma_{\mathrm{SFR}}$-M$_{\star}$ relation we opt to apply some relatively simple metrics and not to over-complicate any analysis beyond what the data would allow.  To this end, we initially simply trace the evolution of the stellar mass point at which the minimum dispersion in the $\sigma_{\mathrm{SFR}}$-M$_{\star}$ relation occurs, hereafter M$^{*}_{x-min}$, where $x$ represents the dispersion metric. The importance of this point to the astrophysical processes occurring in galaxies will be discussed in Section \ref{sec:discuss}. We measure this minimum point for all three of our dispersion metrics and for both the binned data values directly and from our 2$^{nd}$-order polynomial fits.  

In Figure \ref{fig:redevol} we show the evolution of M$^{*}_{x-min}$ for $\sigma_{SD}$ (green points), interquartile range (blue points) and intrinsic scatter (red points) for our 2$^{nd}$-order polynomial fitted values. We also over-plot the $z\sim0$ GAMA measurements from \citep{Davies19a}. Interestingly, we find that for all dispersion metrics we find good agreement in M$^{*}_{x-min}$ at $z<0.55$ and a linearly increasing  M$^{*}_{x-min}$ with redshift. This is also consistent with the point from GAMA at $z\sim0$. These metrics begin to diverge in our 0.55<z<0.7 bin and are inconsistent at the highest redshift bin. 

In order to roughly parameterise the evolution of M$^{*}_{x-min}$, we fit the GAMA and DEVILS M*$_{\sigma_{SD}\mathrm{-min}}$ data values with a linear model (green line). This line is parameterised as:

 \begin{equation}
 \label{eq:minevol}
log_{10}[M^{*}_{x-min}]=1.94(z)+8.86
\end{equation}

While the M$^{*}_{x-min}$ measurements do appear to plateau or drop in the highest redshift bin, potentially where our data does not well-constrain the minimum point, our results do suggest that M$^{*}_{x-min}$ does in fact move to higher stellar masses with redshift/lookback-time. $i.e.$ the point at which scatter along the SFR-M$_{\star}$ relation is smallest occurs at higher stellar masses with redshift. Lastly, we also over-plot the evolution of the `break point' where the star-forming main-sequence flattens at the high mass end from \citet{Thorne21}, defined there as M$_{0}$, for both a fit to the full sample, as presented in their work, and a fit to just the $z>0.45$ data (but extrapolated over all epochs) respectively. We choose to do this as the \citet{Thorne21} work shows a discontinuity in the evolution of M$_{0}$ at $z\sim0.45$, and hence the two reactions given a probable range for the true evolution of M$_{0}$. We find that the evolution of M$_{0}$ is similar to the evolution of M$^{*}_{x-min}$, potentially suggesting a common origin.  The physical interpretation of these trends will be discussed in Section \ref{sec:discuss}.   

\begin{figure*}
\begin{center}
\includegraphics[scale=0.7]{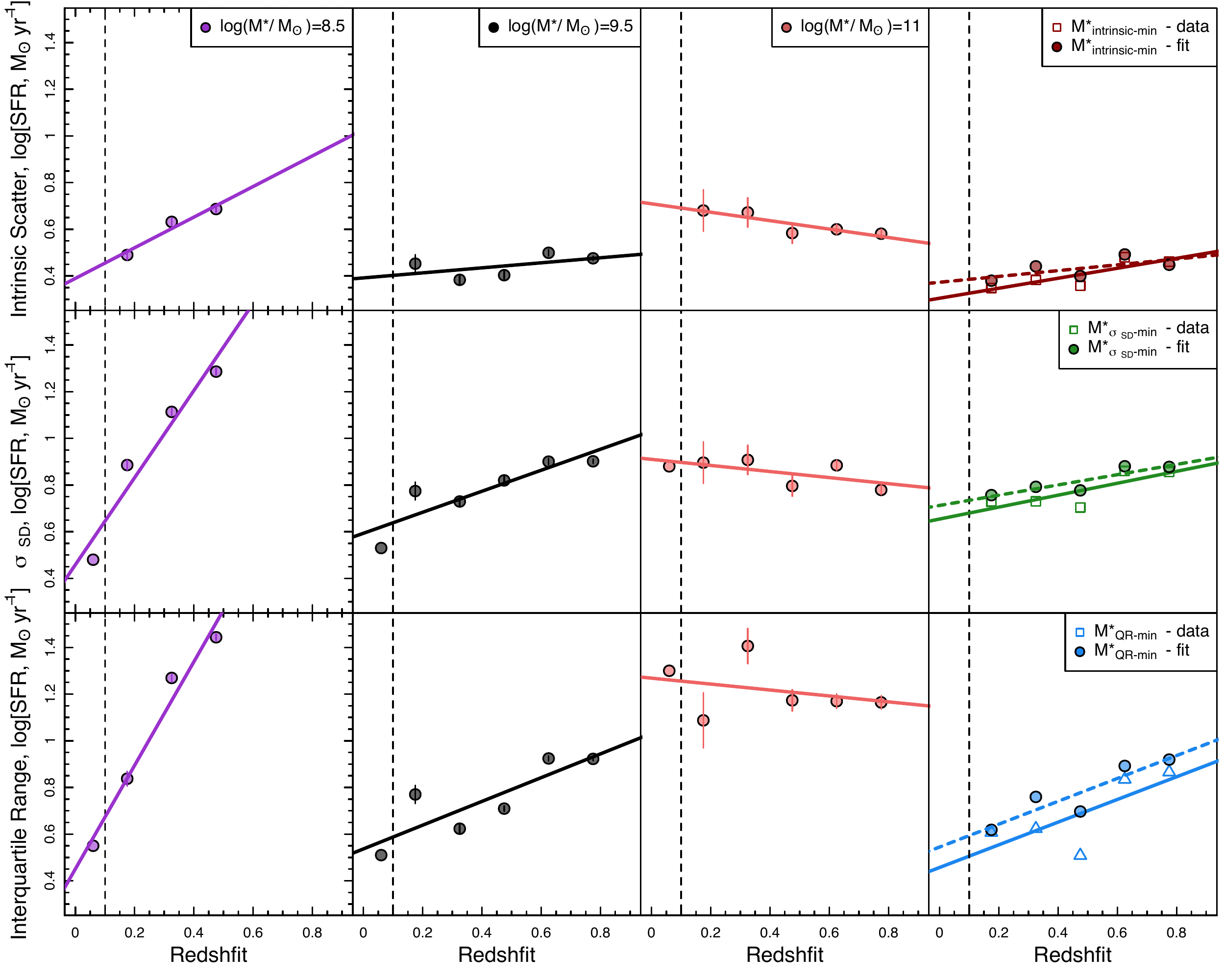}

\caption{Evolution of the dispersion values different four different stellar masses (columns) with redshift using the intrinsic scatter (top row), $\sigma_{SD}$ (middle row), and interquartile range (bottom row). We first show the dispersion at the minimum dispersion point, M*$_{x-min}$, for both the data and our 2$^{nd}$-order polynomial fit as in Figure \ref{fig:redevol}. These stellar mass points bound the region of bellow M*$_{x-min}$ (log$_{10}$[M*/M$_{\odot}$]=8.5), close to M*$_{x-min}$ but at fixed stellar mass (log$_{10}$[M*/M$_{\odot}$]=9.5) and above M*$_{x-min}$ (log$_{10}$[M*/M$_{\odot}$]=11.0). Finally we also over-plot the values for the GAMA sample at $z\sim0$ from \citet{Davies19a} to the left of the dashed vertical line. We fit the evolution of the $\sigma_{SD}$ dispersion at each stellar mass, combining the GAMA and DEVILS results, with a simple linear regression.}
\label{fig:scatevol}
\end{center}
\end{figure*}

\vspace{5mm}

Next we explore the evolution of the shape of the $\sigma_{\mathrm{SFR}}$-M$_{\star}$ relation by plotting the SFR dispersion at three fixed stellar masses:  log$_{10}$[M$_{\star}$/M$_{\odot}$]=8.5, 9.5, 11.0 in the left and middle columns of Figure \ref{fig:scatevol}. These points bound the region of below M$^{*}_{x-min}$ at all epochs, close to M$^{*}_{x-min}$ but at a fixed stellar mass, and above the M$^{*}_{x-min}$, respectively and describe the evolution of the shape of the $\sigma_{SFR}$-M$_{\star}$ relation. Note that for the log$_{10}$[M$_{\star}$/M$_{\odot}$]=8.5 points, we only show results to $z\sim0.5$, as beyond this log$_{10}$[M$_{\star}$/M$_{\odot}$]=8.5 falls below our sample completeness limits and therefore suffers from incompleteness (artificially reducing the dispersion). While the log$_{10}$[M$_{\star}$/M$_{\odot}$]=11.0 points fall above our upper stellar mass limits, we still include them here as our upper limit is somewhat conservative and we wish to compare to previous results, which explore $\sigma_{\mathrm{SFR}}$ at these masses (see below). We then fit the evolution of the minimum dispersion with a linear relation, including the $z\sim0$ GAMA point. 

Within Figure \ref{fig:scatevol}, we find that at the lowest stellar masses the dispersion increases significantly with redshift, suggesting that the scatter about the low stellar mass main-sequence is increasing with lookback-time (which could potentially be evidence of more stochastic processes occurring at higher redshift). Next we find a slight increase in the minimum dispersion at log$_{10}$[M$_{\star}$/M$_{\odot}$]=9.5, which matches the evolution at the M$^{*}_{x-min}$ (as expected).  Finally, we find that the dispersion in the highest stellar mass range is increasing with time. This suggests that the dispersion in the SFR-M$_{\star}$ plane in the most massive galaxies is growing as the Universe evolves. Once again, the physical interpretation of this will be discussed in more details in Section \ref{sec:discuss}.

We then also explore evolution of the measured SFR dispersion value at the M$^{*}_{x-min}$ point for each of our dispersion metrics. The right column of Figure \ref{fig:scatevol} shows the dispersion at M$^{*}_{x-min}$ as a function of redshift for our intrinsic scatter values (dark red points, top panel), $\sigma_{SD}$ (green points, middle panel) and interquartile range (blue points, bottom panel). We then fit the evolution of the minimum dispersion with a linear relation (dark red, green and blue lines respectively). Note that solid lines are fits to the binned data M$^{*}_{x-min}$ points and the dashed lines are fits to our 2$^{nd}$-order polynomial M$^{*}_{x-min}$ measurements - however, these are consistent.  We find that in all three cases the dispersion at M$^{*}_{x-min}$ shows a slight increase with redshift/lookback-time (the solid and dashed dark red, top, green, middle, and blue, bottom lines all slightly increase with redshift). $i.e.$ the minimum dispersion along the SFR-M$_{\star}$ relation gets larger with redshift. Once again, the physical interpretation of this will be discussed in Section \ref{sec:discuss}.

\begin{figure}
\begin{center}
\includegraphics[scale=0.65]{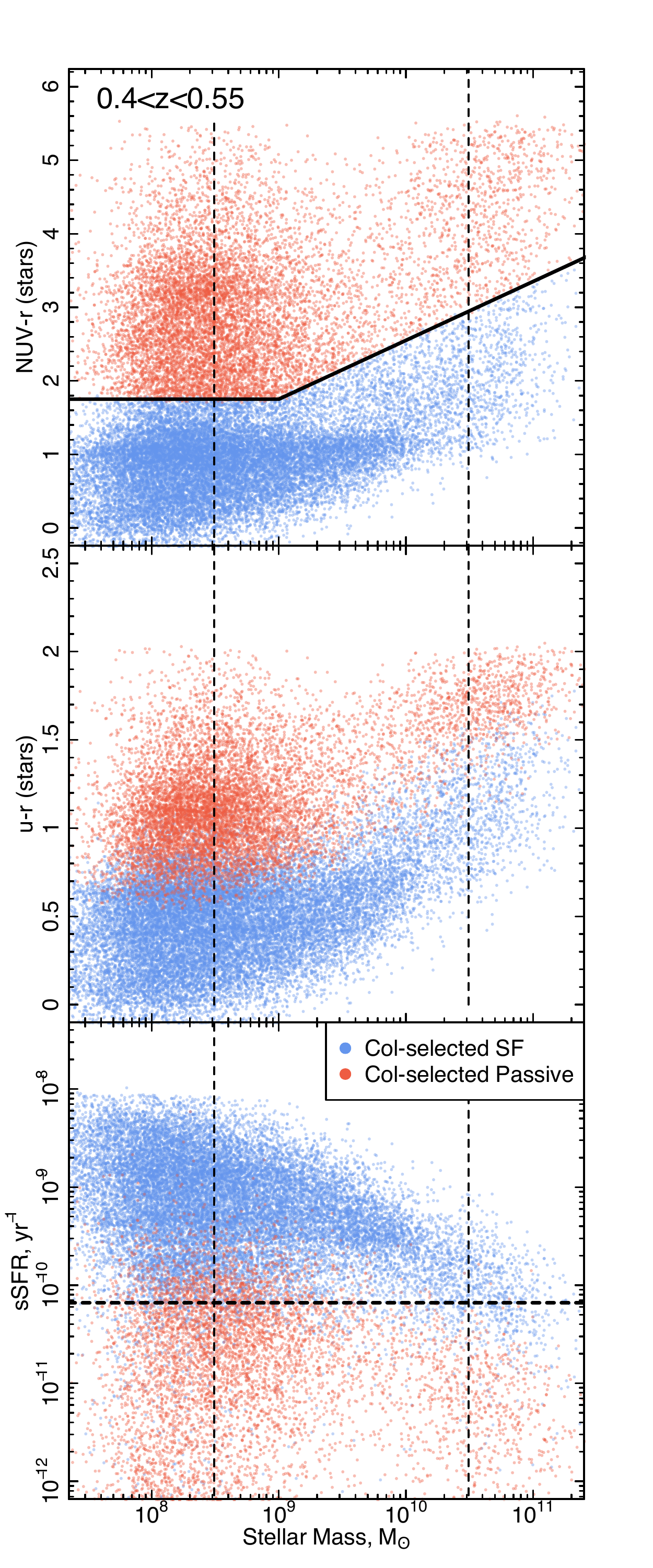}

\caption{The selection of the star-forming population using a NUV-r vs. stellar mass cut in our $0.4<z<0.55$ bin. The top panel displays the selection line given in Equation \ref{eq:colcut}, with the star-forming population in blue and passive population in red. The middle and bottom panels displays the same colour-coding but for rest-frame dust-corrected u-r colour and sSFR, respectively. For reference, in the bottom panel the sSFR cut described in Section \ref{sec:sigma_M_cut} is shown as the dashed horizontal line, highlighting that these selections are roughly comparable. Once again, dashed vertical lines display our sample fitting ranges at this epoch.}
\label{fig:colourSel}
\end{center}
\end{figure}

\begin{figure*}
\begin{center}
\includegraphics[scale=0.6]{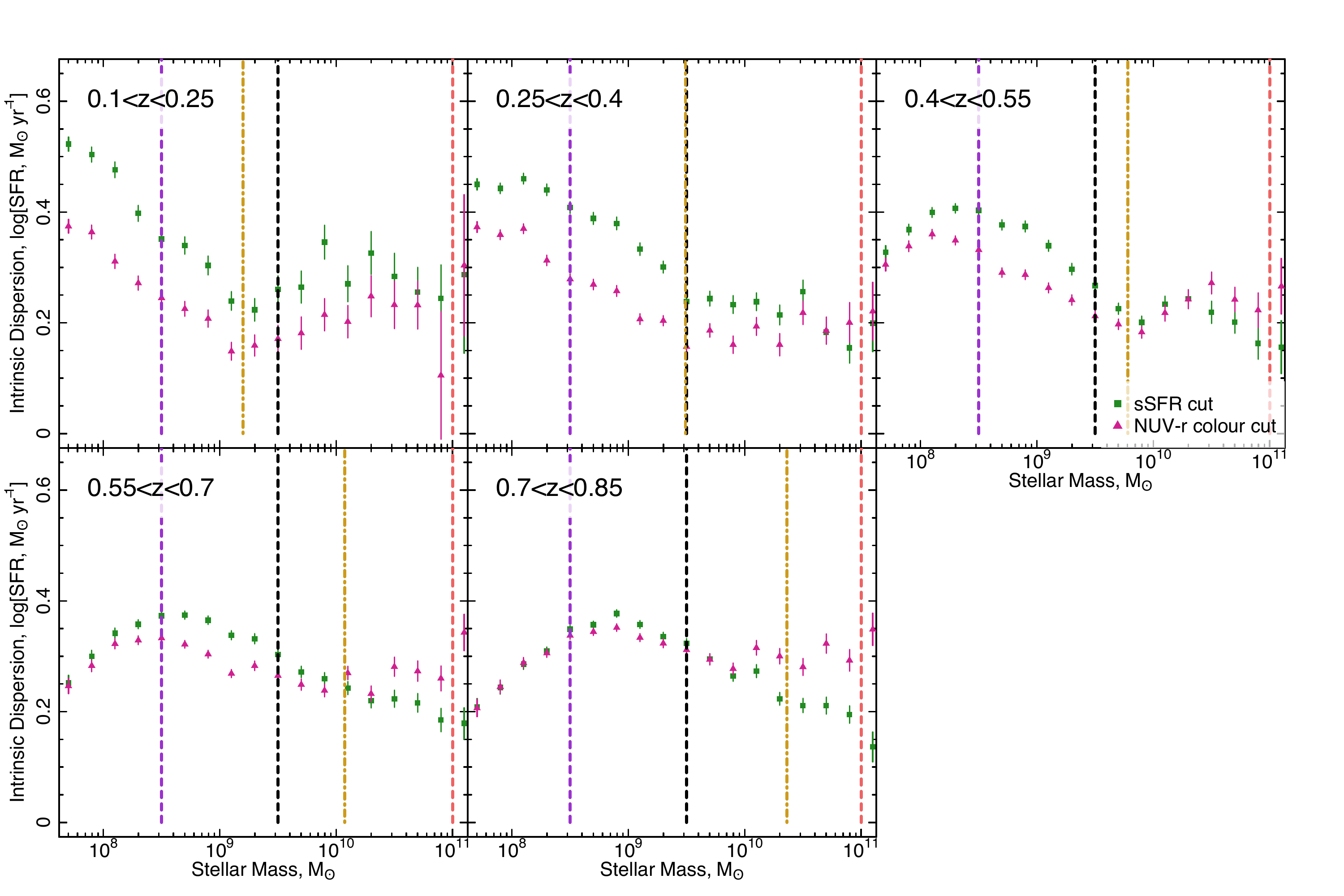}

\caption{The same at Figure \ref{fig:Msig} but just showing the intrinsic dispersion measurements with a cut of sSFR$>$MS$_{9-10}$ - 1\,dex applied (green points) and a NUV-r colour-selection (violet points) applied, to be directly comparable to previous observations and simulation results, which use similar cuts. We note here that the dashed vertical lines do not display stellar mass ranges, as in Figure \ref{fig:Msig}, but reference stellar mass points for comparison. Some of the previous literature works compare the evolution of the dispersion at a number of fixed stellar mass points. We show these here as vertical purple, black and red dashed lines at log$_{10}$[M$_{\star}$/M$_{\odot}$]=8.5, 9.5 and 11 respectively. The vertical dot-dashed gold line also shows the parameterisation of the evolution of M$^{*}_{x-min}$ from Equation \ref{eq:minevol}. Visually, this is close to the minimum dispersion point in the colour selected sample for all epochs at $z<0.7$.}
\label{fig:Msig_cut}
\end{center}
\end{figure*}

\subsection{The $\sigma_{\mathrm{SFR}}$-M$_{\star}$ relation for just the SFS} 
\label{sec:sigma_M_cut}

Comparing our results with previous measurements of the evolution of the $\sigma_{\mathrm{SFR}}$, both from observations and simulations, is somewhat complicated. This is largely due to the fact that the majority of these studies quote the dispersion along the star-forming sequence only (excluding passive systems). We opt not to do this in the previous sections as the methods by which the passive systems are removed are varied and sometimes disparate, and the choice of this method can have a strong impact on the measured dispersion at particular stellar masses \citep[$i.e.$ see][]{Davies19b}. This coupled with the fact that these previous works use different SFR indicators, different methods for determining redshifts and stellar masses, have different selection limits,  makes comparing $\sigma_{\mathrm{SFR}}$ measurements fraught with difficulty. Thats said, we do aim to provide such a comparison here in order to roughly place our new measurements within the context of existing works. 

\vspace{2mm}

In order to directly compare to these samples, we first must isolate the star-forming main-sequence. To do this, we opt for two simple methods which are comparable to those  is used in many of the previous studies, first using a simple sSFR cut to exclude the passive population and second using rest-frame colours to identify blue star-forming galaxies. At each redshift bin in our sample, for the former we determine the normalisation of the star-forming main-sequence at 9$<$log$_{10}$[M*/M$_{\odot}$]$<$10 in terms of sSFR (hereafter, MS$_{9-10}$) and simply deem all galaxies that have sSFRs of greater than 1\,dex below MS$_{9-10}$ as star-forming. While this selection is somewhat arbitrary, it is largely consistent with many of the previous works (see the appendix). An example of this selection line at $0.4<z<0.55$ is shown in Figure \ref{fig:M-SFR}. Next, we perform a colour-selection to isolate the star-forming population following a similar method to \cite{Ilbert15} and using the rest-frame, dust-corrected photometry derived in the \textsc{ProSpect} analysis of \cite{Thorne21}. We initially visually define a dividing line between the star-forming and passive populations in the NUV-r vs. stellar mass plane for our $z\sim0$ population, and then apply a similar normalisation offset with lookback time as \cite{Ilbert15}, giving a star-forming selection of:

\begin{equation}
\label{eq:colcut}
NUV^{*}-r^{*}>2 - 0.09(t_{\mathrm{lb}}-2.21) \,\,\, [M^*<10^{9}\,M_{\odot}] 
\end{equation}   
\begin{equation*}
NUV^{*}-r^{*}>0.8\mathrm{log}_{10}[M^*/M_{\odot}] - 5.2 - 0.09(t_{\mathrm{lb}}-2.21) \,\,\, [M^*>10^{9}\,M_{\odot}]
\end{equation*} 

\noindent where NUV$^{*}$-r$^{*}$ is the rest-frame, dust-corrected NUV-r colour and $t_{\mathrm{lb}}$ is the lookback time in Gyr. The 0.09($t_{\mathrm{lb}}$-2.21) scaling factor is adapted from \cite{Ilbert15} taking into account the changing colours of galaxies with redshift. This selection is displayed in Figure \ref{fig:colourSel} for our $0.4<z<0.55$ sample. Here we show our selection plane in the top panel and points selected as passive and star-forming in the u-r vs. stellar mass plane (middle) and sSFR vs stellar mass (bottom). The bottom panel indicates that this selection is comparable to the sSFR selection at this epoch, but does not provide a hard cut in sSFR that could impact the measured dispersion values.       

Using the isolated SFS population, we then repeat our measurements of the $\sigma_{\mathrm{SFR}}$-M$_{\star}$ relation for each of our dispersion metrics. Figure \ref{fig:Msig_cut} shows the intrinsic scatter for both star-forming population selections at each redshift. Note that we do not show $\sigma_{SD}$ and interquartile range dispersions here to reduce complexity. However they do show similar trends with stellar mass but with larger dispersion normalisation (as in Figure \ref{fig:Msig}). From this figure we see that (as expected) at low stellar masses the dispersion along the main-sequence is largely the same as our previous results (where the bulk of the galaxies are not passive), but  significantly change at the high stellar mass end, where a large fraction of the systems are removed - particularly when using a sSFR cut. We also find that using either a sSFR or colour selection yields similar results at low/intermediate stellar masses in shape (with some normalisation offset), but differ slightly at the high stellar mass end. For the colour-selection the `U-shaped' distribution can still been seen (albeit more weakly), while in the harsher sSFR-selection this shape is removed in all but the lowest redshift bin, and the intrinsic scatter declines with stellar mass. This is consistent \cite{Davies19a} for the $z\sim0$ population, and some previous studies (see below and the Appendix), where a linear sSFR cut largely removes some of the scatter in the high stellar mass population. However, as the `U-shaped' distribution remains when using a colour selection of the SFS, it is likely not simply due to the passive population, but also caused by an increased dispersion within the main-sequence at the high stellar mass end.      

\begin{figure*}
\begin{center}
\includegraphics[scale=0.47]{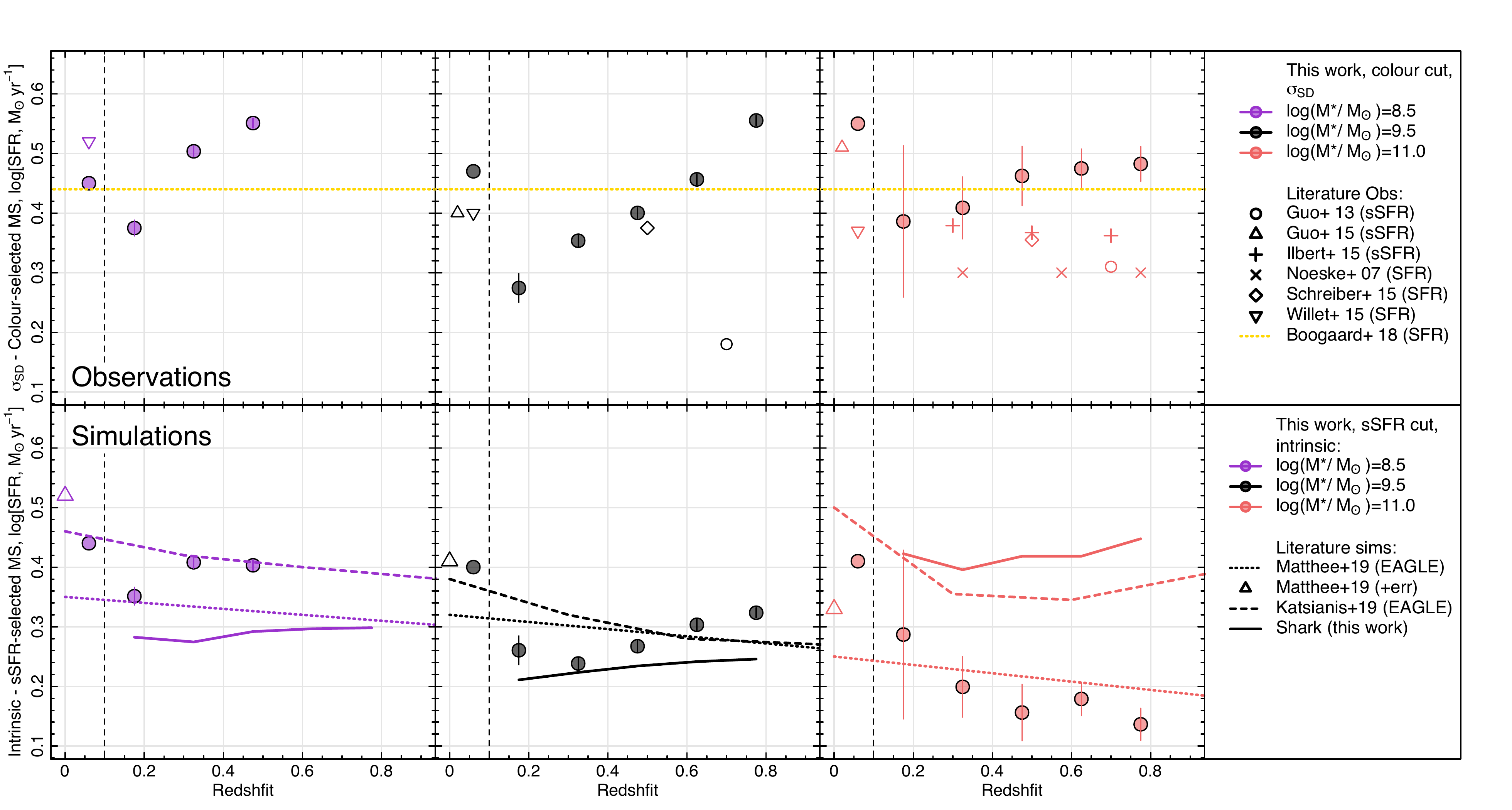}
\caption{Evolution of the SFR dispersion values at different stellar masses with redshift when a sSFR cut of sSFR$>$(MS$_{9-10}$ - 1\,dex) is applied (top row) and a NUV-r colour cut is applied (bottom). Here we compare to previous observational measurements (top row) and simulation predictions (bottom row) - given in the right hand legend. We opt to display the colour-selection with $\sigma_{SD}$ dispersion in the top row and sSFR selection with intrinsic dispersion in the bottom row as these most closely match the selections applied in the literature observations and simulations, respectively. We use the same stellar mass points as \citet{Katsianis19} to be able to directly compare to the EAGLE predictions. We also over-plot the values for the GAMA sample at z$\sim0$ from \citet{Davies19a} - again to the left of the dashed vertical line. Literature data points are colour-coded by the closest match to our three stellar mass bins.}
\label{fig:scatevolComp}
\end{center}
\end{figure*}

We present a detailed comparison to previous observations \citep{Noeske07, Guo13,Guo15,Ilbert15,Schreiber15,Willett15,Boogaard18}, the EAGLE hydrodynamical simulations \citep{Matthee18,Katsianis19} and Shark Semi analytic model \citep{Lagos18} in the Appendix, which is summarised in Figure \ref{fig:scatevolComp}. However, to give an overview of these comparisons here, in terms of the previous observational results, we see that the observational picture is complicated and varied. However, despite the samples being selected in very different ways and the $\sigma_{\mathrm{SFR}}$-M$_{\star}$ relation measured with varied approaches, our new measurements are largely consistent with the existing literature. They have the roughly the same $\sigma_{SD}$ values for high stellar mass galaxies, and largely the same evolutionary trend. The only exception to this is the log$_{10}$[M$_{\star}$/M$_{\odot}$]$=$9.5 point from \cite{Guo13} at $z=0.7$, which is in strong contention with our new measurements. However, we note that this measurement is made at a point where the majority of their sample is undetected in the observational band used to measure their SFRs. 

In terms of the EAGLE results, we note that despite using the same simulation \cite{Matthee18} and \cite{Katsianis19} find distinctly different results, highlighting how the choice of SFS selection and methodology can strongly affect the measurement of the $\sigma_{\mathrm{SFR}}$-M$_{\star}$ relation. However, to first order none of the EAGLE predictions are in strong contention with our observational trends.  This is particularly true for \cite{Matthee18} who find a marginally decreasing intrinsic $\sigma$ at all stellar masses and decreasing intrinsic $\sigma$ with stellar mass, as our observations. The \cite{Katsianis19} results are similar for our low and intermediate stellar mass samples, but they find a much larger intrinsic $\sigma$ for high stellar mass galaxies. The $\sigma$ measurement for high stellar mass galaxies is most sensitive to choice of sSFR, and thus is very dependant on exact methodology, so this is unsurprising.  In addition, it is a these masses that the most tension exists between the results of \cite{Matthee18} and \cite{Katsianis19}.

For Shark, we can apply the exact same methodology as for DEVILS (see appendix \ref{sec:shark} and Figure \ref{fig:SharkCompApp}), including the same measurement of intrinsic dispersion and the same sSFR cuts. Exploring the evolution of the dispersion at our three stellar mass points (Figure \ref{fig:scatevol}), we see very little evolution at any stellar mass and the results differ from both our observational results and EAGLE, in that even with a sSFR selection the largest SFR dispersion in the Shark sample occurs at the highest stellar masses. This largely occurs as using a sSFR cut defined at 1 dex below the SFS still retains a significant fraction of passive galaxies in the Shark sample. However, for direct comparison, we do not change the method which we use for the observational data. We do note, that the Shark dispersion close to M$^{*}_{x-min}$ ($i.e.$ at log$_{10}$[M$_{\star}$/M$_{\odot}$]=9.5) lies close to our measured observational values when a sSFR selection is applied (black points compared to lowest blue line in the bottom panel of Figure \ref{fig:scatevol}). We therefore, once again highlight that the Shark SFS minimum dispersion values are close to our observational data, but the evolution of the shape of the $\sigma_{\mathrm{SFR}}$-M$_{\star}$ relation is not consistent. For a much more detailed comparison with these previous results, please see the Appendix.

\subsection{Two SFS populations split at M$^{*}_{x-min}$?} 
\label{sec:distributions}

\begin{figure*}
\begin{center}
\includegraphics[scale=0.55]{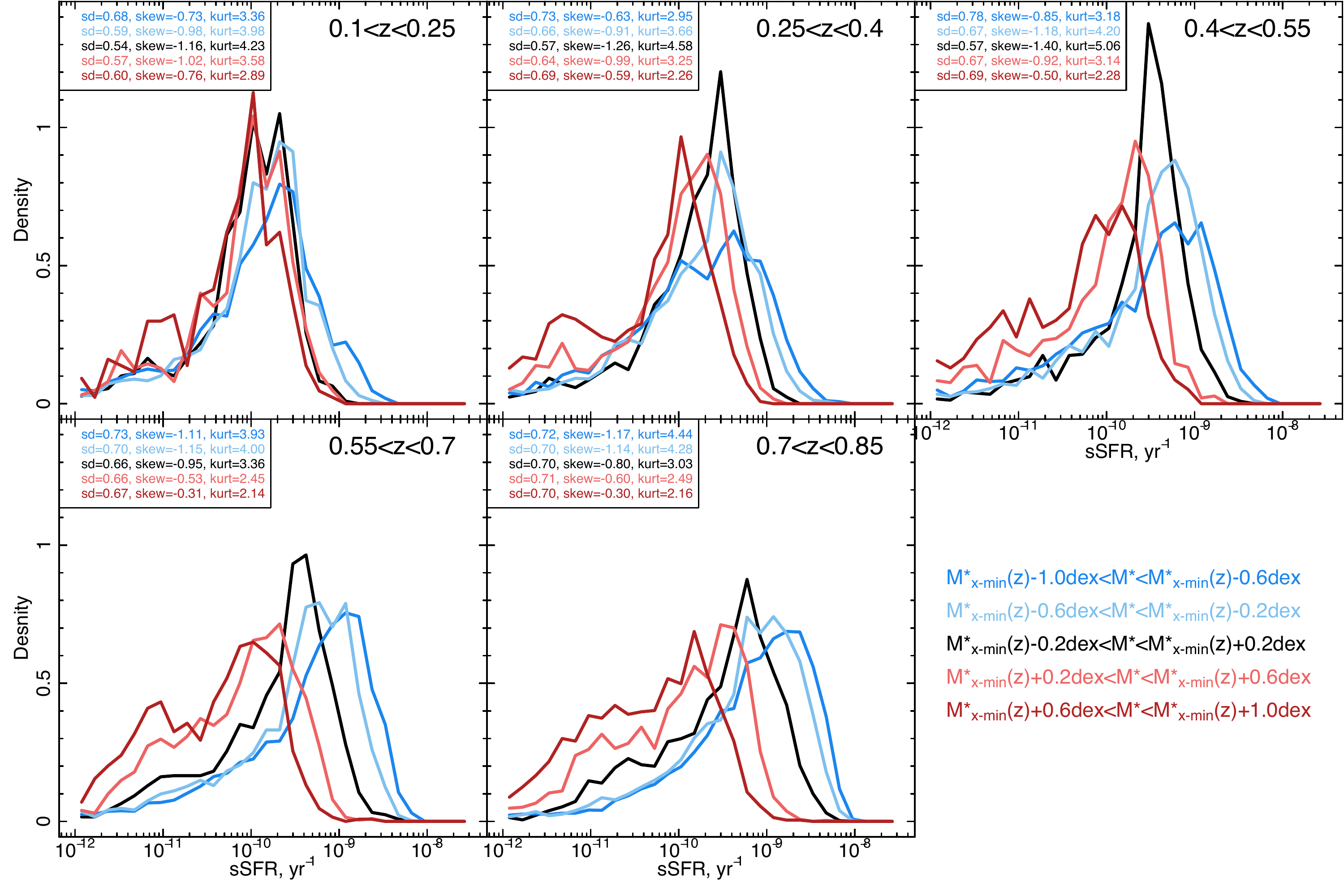}


\includegraphics[scale=0.55]{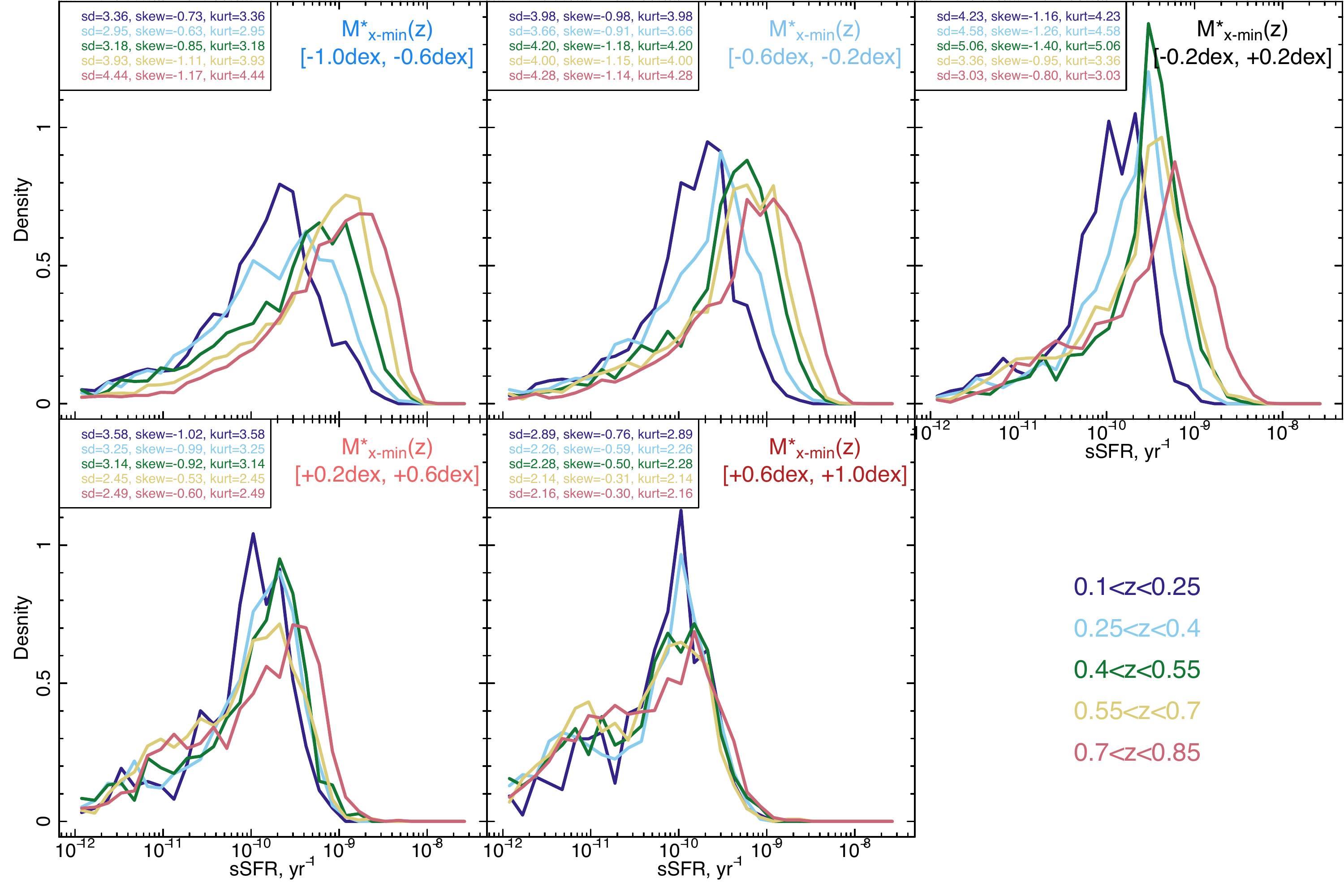}

\caption{The distribution of sSFRs as a function of stellar mass with respect to M$^{*}_{x-min}$($z$) at a given redshift (top panels) and the distribution of sSFRs as a function of redshift at a given stellar mass, with respect to M$^{*}_{x-min}$ ($z$) (bottom panels). The standard deviation, skewness and kurtosis for each distribution is given in the legend.}
\label{fig:dists}
\end{center}
\end{figure*}

Following our analysis of the $\sigma_{\mathrm{SFR}}$-M$_{\star}$ relation it is also informative to consider the full distribution of the sSFRs across the sSFR-M$_{\star}$ plane, and it is evolution. For this we split our sample at each redshift into five stellar mass bins, which are defined relative to M$^{*}_{x-min}$ at a given epoch. Two of these cover regions below M$^{*}_{x-min}$, one at M$^{*}_{x-min}$, and two above M$^{*}_{x-min}$. In the top panels of Figure \ref{fig:dists} we then display the distribution of sSFRs in each of these stellar mass ranges, with each panel showing a different epoch. We also calculate the standard deviation, skewness and kurtosis of each distribution (in log$_{10}$[sSFR]), given in the legend of each panel.  Taking each panel independently, we see that if we compare the low mass (blue) to high mass (red) distributions we find that they i) shift to lower sSFRs, ii) become broader, iii) typically have lower kurtosis and iv) appear more bimodal. This is generally all expected as we see more red and passive galaxies at higher stellar masses, observed as an additional population broadening the distribution in the sSFR-M$_{\star}$ plane. We find that the highest kurtosis values occur at M$^{*}_{x-min}$ at all epochs, once again indicating that the distribution is most strongly peaked at the minimum dispersion point.  We also find that the diversity/spread of sSFRs appears to increase with redshift ($i.e.$ the full spread of galaxy sSFRs is broader), this is consistent with the $\sigma_{\mathrm{SFR}}$-M$_{\star}$ relation increasing with normalisation and is indicative of more turbulent star formation in the earlier Universe.  

In the bottom panels of Figure \ref{fig:dists} we then compare the evolution of the sSFRs distributions at fixed stellar mass with respect to M$^{*}_{x-min}$. What is most striking in these panels is that there is very little evolution in the distribution of sSFRs for high mass galaxies in terms of median sSFR, but strong evolution in low stellar mass galaxies. This suggests that over the last $\sim6$\,Gyr the high mass end of the SFS has evolved very little in sSFR, while the low mass end evolves considerably. We reiterate that these stellar mass ranges are defined with respect to the M$^{*}_{x-min}$ point. As such, this leads to an intriguing possibility, is the SFS evolving differently above and below M$^{*}_{x-min}$? \cite{Thorne21} show that for the DEVILS sample used here there is a turnover stellar mass in the SFS that increases to higher stellar masses with lookback time (defined at $M_{0}$). As shown in Figure \ref{fig:redevol} this point lies close to M$^{*}_{x-min}$, and using two different methods for fitting $M_{0}$, its value bounds  M$^{*}_{x-min}$ at all epochs. This, combined with the indication different evolution of sSFRs above and below M$^{*}_{x-min}$, leads to the question of whether M$^{*}_{x-min}$ (the minimum dispersion point on the SFS) also traces the break point in the SFS.        

\begin{figure*}
\begin{center}
\includegraphics[scale=0.58]{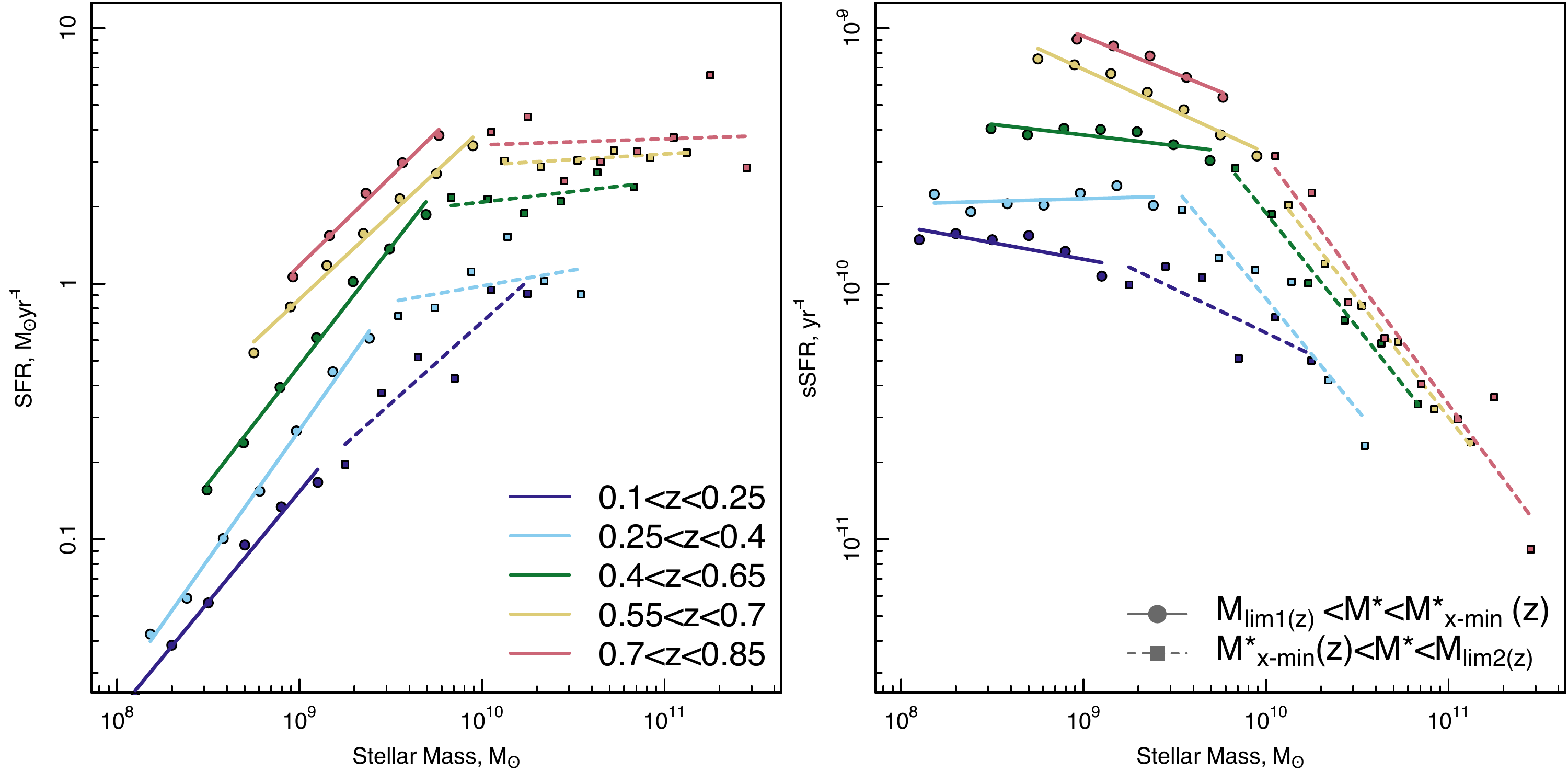}
\caption{Fits to the median SFS as a function of redshift split at M$^{*}_{x-min}$($z$) in terms of SFR (left) and sSFR(right). Different coloured lines show the various epoch explored, while circles and solid lines show the median and fit below M$^{*}_{x-min}$($z$), and squares and dashed lines show median and fit above. M$^{*}_{x-min}$($z$)  The M$^{*}_{x-min}$($z$) point does appear to trace the break in the SFS at all epochs. Below M$^{*}_{x-min}$($z$) we find a strong slope in the SFR-M$_{\star}$ distribution, but above M$^{*}_{x-min}$($z$) the SFR-M$_{\star}$ is flat ($i.e.$ all galaxies have a fixed SFR irrespective of stellar mass). We also find that below M$^{*}_{x-min}$($z$) we see stronger evolution in both the slope and normalisation of the sequence than above M$^{*}_{x-min}$($z$). This is true for all but the lowest redshift bin. However, \citet{Thorne21} show that at this epoch the DEVILS sample alone does not contain enough high stellar mass galaxies to accurately constrain the break in the SFS.}
\label{fig:SFSevol}
\end{center}
\end{figure*}

To explore this further, Figure \ref{fig:SFSevol} displays the SFS evolution, where we split the sample above and below the M$^{*}_{x-min}$ point at a given epoch. For ease of description, we display the SFS both in the SFR-M$_{\star}$ (left) and sSFR-M$_{\star}$ (right) planes. To fit these relations we simply take the median SFR/sSFR as a function of stellar mass for all sSFR>$10^{-12}$M$_{\odot}$ galaxies and then fit a linear relation to the median points. Interestingly, this figure shows exactly what was predicted above. The M$^{*}_{x-min}$ does in fact appear to trace the break in the SFS at all epochs. We do also see distinctly different properties of the SFS either side of M$^{*}_{x-min}$. Below this point the SFS is steep in terms of SFR ($\sim$flat in sSFR) and evolves strongly in both slope and normalisation. However, above M$^{*}_{x-min}$ the SFS is flat in terms of SFR (slope of 1 in terms of sSFR), does not evolve in slope, and evolves less strongly in normalization. 

In combination, these results suggest that the M$^{*}_{x-min}$ point, not only traces the minimum dispersion point along the SFS, but also traces the break point in the SFS, where the sequence evolves differently either side of M$^{*}_{x-min}$. Potentially this indicates that the M$^{*}_{x-min}$ point delineates the boundary point between two different evolutionary mechanisms that are diving star-formation changes in galaxies.

\begin{figure*}
\begin{center}
\includegraphics[scale=0.38]{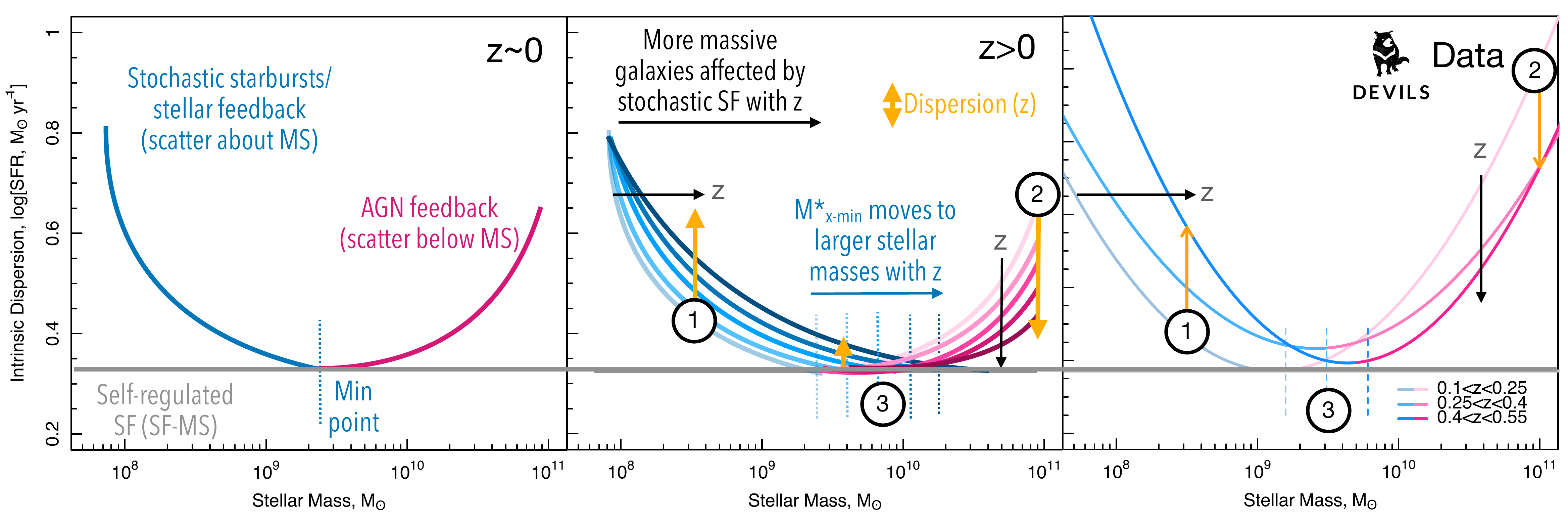}

\caption{Infographic showing the shape of the physical interpretation of the $\sigma_{\mathrm{SFR}}$-M$_{\star}$ relation and its evolution with redshift. The left panel shows the physical interpretation of the shape of the $\sigma_{\mathrm{SFR}}$-M$_{\star}$ relation at a single epoch. In the middle we show a representation of how the $\sigma_{\mathrm{SFR}}$-M$_{\star}$ relation evolves with time and highlight the key observable trends numbered 1-3, see Section \ref{sec:discuss} for further discussion of this infographic's key components. On the right we show the fits the $intrinsic$ scatter for our DEVILS data at $z<0.55$ (taken from Figure \ref{fig:M-SFR}) plotted in an identical way to our infographic and showing the same trends.}
\label{fig:cartoon}
\end{center}
\end{figure*}

\section{Towards a Physical Interpretation of the $\sigma_{\mathrm{SFR}}$-M$_{\star}$ relation and its Evolution}
\label{sec:discuss} 

In order to form a physical interpretation of our results, we first summarise the key observational trends discussed in the previous section:\\

\noindent $\bullet$  When no cuts are applied to the DEVILS sample we find a `U-shaped'  $\sigma_{\mathrm{SFR}}$-M$_{\star}$ with high dispersion at the low and high stellar mass end, and minimum dispersion point at around 9$<$log$_{10}$[M*/M$_{\odot}$]$<$10. This is true when using either the $\sigma_{SD}$, interquartile range or the intrinsic scatter as a measure of dispersion, and is consistent with GAMA at $z\sim0$ (Figure \ref{fig:Msig}). \\

\noindent $\bullet$  The minimum dispersion point, M$^{*}_{x-min}$, appears to evolve $\sim$linearly with redshift to $z\sim0.8$, increasing in stellar mass. This is true to $z\sim0.6$ using either the data directly, or a 2$^{nd}$ order polynomial fit, and for $\sigma_{SD}$, interquartile range and intrinsic scatter (Figure \ref{fig:redevol}).  \\

\noindent $\bullet$  When no cuts are applied to the DEVILS sample the dispersion at M$^{*}_{x-min}$ increases with redshift from $0<z<0.8$. This is true using either the data directly, or a 2$^{nd}$ order polynomial fit, and for all dispersion measurements. At low stellar masses the dispersion increases significantly at $0<z<0.5$ (beyond this our sample is incomplete at these masses) and at high stellar masses the dispersion decreases (Figure \ref{fig:scatevol}). \\

\noindent $\bullet$ When we apply a sSFR cut  to isolate the SFS the `U-shaped' distribution is largely removed (in all but the lowest redshift bin) and we find the $\sigma_{\mathrm{SFR}}$ decreases with stellar mass. Applying a NUV-r colour selection (Figure \ref{fig:colourSel}) the `U-shape' is still present but weakened and has a minimum that is consistent with M$^{*}_{x-min}$ for the full sample (Figure \ref{fig:Msig_cut}).\\

\noindent $\bullet$ For a NUV-r colour selection and $\sigma_{SD}$ we find that the dispersion increases with redshift at low stellar masses (log$_{10}$[M*/M$_{\odot}$]$=$8.5) and intermediate stellar masses (log$_{10}$[M*/M$_{\odot}$]$=$9.5) and is roughly flat with redshift at high stellar masses (log$_{10}$[M*/M$_{\odot}$]$=$11). Literature observations only really have measurements at the high stellar mass end at $z>0$, and these are largely consistent with our measurements at log$_{10}$[M$_{\star}$/M$_{\odot}$]$=$11, but with small normalisation offsets (Figure \ref{fig:scatevol} - top panel). \\

\noindent $\bullet$ For a NUV-r colour selection and intrinsic dispersion we find that $\sigma_{\mathrm{SFR}}$ marginally decreases or is flat with redshift at all stellar masses. The EAGLE simulations also find that $\sigma_{\mathrm{SFR}}$ deceases with redshift at all stellar masses, modulo slightly different results from two different EAGLE studies (Figure \ref{fig:scatevol} - bottom panel). \\

\noindent $\bullet$ While the Shark semi-analytic model shows similar intrinsic SFR dispersion values to our DEVILS observations, it does not show significant evolution in the shape of the $\sigma_{\mathrm{SFR}}$-M$_{\star}$ relation, and therefore also M$^{*}_{x-min}$ (Figures \ref{fig:redevol} and \ref{fig:scatevol} - bottom panel). \\

\noindent $\bullet$ If we isolate the SFS above and below the M$^{*}_{x-min}$ point, we find that it appears to trace the break in the SFS, and we observe different evolution of the SFS in these two regimes.  \\

Next using these observations we propose potential physical interpretations which could lead to these observational trends. First, we reiterate the previously proposed physical interpretations of the `U-shape' of the $\sigma_{\mathrm{SFR}}$-M$_{\star}$ at $z\sim0$ and revisit this interpretation in the context of our new results.

In this model the dispersion is high at the low stellar mass end due to stochastic star-bursts and stellar feedback events leading to a $log-normal$ symmetrical `puffing-up' of the SFS, while at the high stellar mass end quenching events, such as feedback from AGN, cause galaxies to drop off the SFS and become passive increasing the dispersion and making the distribution more asymmetric or bimodal. At intermediate stellar masses, galaxies are too massive to be strongly impacted by stochastic feedback processes and not massive enough to have grown strong, powerful AGN which can cause galaxy-wide quenching. As such, they exist in a self-regulated star-formation state where there is a ready supply of star-forming gas and galaxies grow uniformly. This results in a tight, low-dispersion SFS. It must be noted here, that this assumption is likely only true for central galaxies. Satellite galaxies can undergo additional environmental quenching mechanisms which lead to increased dispersion in the SFR-M$_{\star}$ plane. This dispersion is likely to occur primarily at intermediate stellar masses, impacting the overall shape of the distribution and likely removing the `U-shaped' dispersion \citep[$i.e.$ see][]{Davies19a}. As we still clearly see the `U-shape' in our results here, and environmental impact potentially removes this shape, this may suggest that we are not strongly impacted by environment. However, in this work we choose not to consider the impact of environmental quenching and/or central/satellite status as these data products do not currently exist for the DEVILS sample. Instead, we defer further analysis in this area to subsequent DEVILS papers.

This model for the physical interpretation of the shape of the  $\sigma_{\mathrm{SFR}}$-M$_{\star}$ relation is described and justified in detail in \cite{Katsianis19} and \cite{Davies19a}. In \cite{Katsianis19} they show that this model is consistent with EAGLE, as running the simulation with AGN feedback turned off reduces the dispersion at the high stellar mass end, while suppressing stellar feedback reduces the scatter at the low stellar mass end. As such, in our interpretation here we first start from the premise that these assumptions regarding the physical origin of the high dispersion at each end of the SFS are correct, and then aim to explain the evolutionary trends we observe in our data. We note that this physical interpretation is not necessarily the true astrophysical mechanisms which are occuring, and we will discuss different potential interpretations and caveats throughout this section. 

To start, the left panel of Figure \ref{fig:cartoon} displays an infographic of the model described above. The grey line represents a tight, uniform dispersion main sequence with self-regulated star-formation and ready supply of gas. Adding in stochastic processes that lead to starburst and (short-duration) quenching events leads to an upturn in the dispersion at the low mass end (blue line), while adding passive systems and large-scale AGN quenching leads to an upturn in the dispersion at the high mass end. In the middle panel of Figure \ref{fig:cartoon} we then aim to visually represent the three key observational trends in our data that we believe to be representative of the astrophysical processes that are shaping the evolution of $\sigma_{\mathrm{SFR}}$-M$_{\star}$:\\
\\

\noindent $\bullet$ 1. Below M$^{*}_{x-min}$ the dispersion of SFRs seen in galaxies at a fixed stellar mass increases with redshift. Assuming this dispersion is produced by stochastic stellar processes, this would suggest that these processes are are more extreme at earlier times and can produce more significant starburst and/or feedback events at a fixed stellar mass. This is consistent with our understanding of galaxy evolution processes, where sSFRs at a fixed stellar mass are larger, gas rich merger events more prevalent and starbursts more extreme (due to available gas supply and lower metallicity populations) at earlier times. As such, galaxies which have low enough stellar mass to be globally affected by both single starburst/feedback events, will be significantly driven off the SFS leading to increased dispersion. However, it must be noted here that this does not provide evidence for or against stochastic stellar processes being the cause of increased dispersion at low stellar masses, only that our observations are consistent with this assumption. The increased scatter could equally be produced by a larger variation in longer-duration SFHs at lower stellar masses. Exploring this will be the  subject of a follow-up paper (Davies et al in prep). \\

\noindent $\bullet$ 2. Above M$^{*}_{x-min}$ the dispersion of SFRs seen in galaxies at a fixed stellar mass decreases with redshift. At the high stellar mass end, this dispersion measurement is largely driven by the (permanently) quenched population combined with actively quenching galaxies which produce asymmetric scatter to low SFRs, $i.e.$ there are very few star-bursting massive galaxies. As the passive population grows with time, it is unsurprising that the dispersion in SFRs increases as more and more galaxies fall to below the main sequence. This is observed for our high mass populations in DEVILS. We can not, with our current results, make any statement regarding the origin of the high stellar mass scatter ($i.e.$ if it is caused by AGN feedback) but will also explore the distribution of AGN across this plane in Davies et al (in prep) to help elucidate this picture. \\

\noindent $\bullet$ 3. The M$^{*}_{x-min}$ point moves to larger stellar masses with redshift and the dispersion measurement at M$^{*}_{x-min}$ slightly increases. From the discussion in point 1, we suggest that the ability of starburst/feedback events to strongly impact the position of a galaxy within the SFR-M$_{\star}$ plane at a fixed stellar mass increases with lookback time. The likelihood of these event are largely driven by two properties, the sSFR (secular starbursts) and gas rich major-mergers. We also know that at a fixed stellar mass, the sSFR of galaxies increases with lookback time, and the incidence of major mergers at a fixed stellar mass increases with lookback time. Following this, it is likely that the stochastic processes leading to large dispersion in SFRs, can occur in more and more massive galaxies as we look further back into the early Universe - $i.e.$ the energy input/output from a system from star-formation per unit stellar mass can get larger with lookback time. This naturally leads to a scenario with the M$^{*}_{x-min}$ point (the point where stochastic processes can begin to cause large dispersion) moves to higher stellar masses with redshift - as seen in our observations.\\

Following this logic and being speculative, the M$^{*}_{x-min}$ point, and its evolution can be thought to represent both the minimum stellar mass where a galaxy is not significantly affected by stellar processes leading to starburst/quenching, and at earlier times the maximum stellar mass where a galaxy is not able to be made passive by large-scale quenching process ($i.e.$ AGN feedback). Therefore, M$^{*}_{x-min}$ represents the point in the SFR-M$_{\star}$ plane where galaxies exist in a stable, self-regulated star-formation regime - $i.e.$ the balance point in galaxy feedback processes. As such, it is likely a highly important property in understanding the global evolution of galaxies and the processes by which they move within that SFR-M$_{\star}$ plane - eventually becoming quenched. 

Interestingly, the stellar masses at which a galaxy is quenched by large-scale processes ($i.e.$ AGN feedback) is potentially more stable in evolution than the point at which star-formation processes can effect a galaxy's position with respect to the SFS. For example, the former is driven by AGN feedback, which is correlated with a galaxy's black-hole mass and therefore bulge mass (via the M$_{BH}$-$\sigma_{v-bulge}$ relation). Within this regime, to first order, the energy input into the system from AGN leading to quenching is directly correlated with the stellar mass of the galaxy, and the property resisting the quenching ($i.e.$ largely gravity to retain star-forming gas) is also correlated with stellar mass, leading to self-similar evolution. However, the latter is governed by SFR which, at a fixed stellar mass, decrease strongly with time. This means that as the Universe evolves less energy per unit stellar mass is input into galaxies via star-formation processes, reducing the likelihood of strong starburst/feedback events at a given stellar mass. Thus, the upturn in the dispersion in SFR at low stellar masses will likely evolve more strongly than the upturn at the high mass end. This is seen somewhat in our Figure \ref{fig:scatevol}, but will require further study with larger samples extending to lower stellar mass limits - $i.e.$ the combination of 4MOST-WAVES-deep \citep{Driver19} and deep LSST+Nancy Grace Roman imaging.

\begin{figure*}
\begin{center}
\includegraphics[scale=0.8]{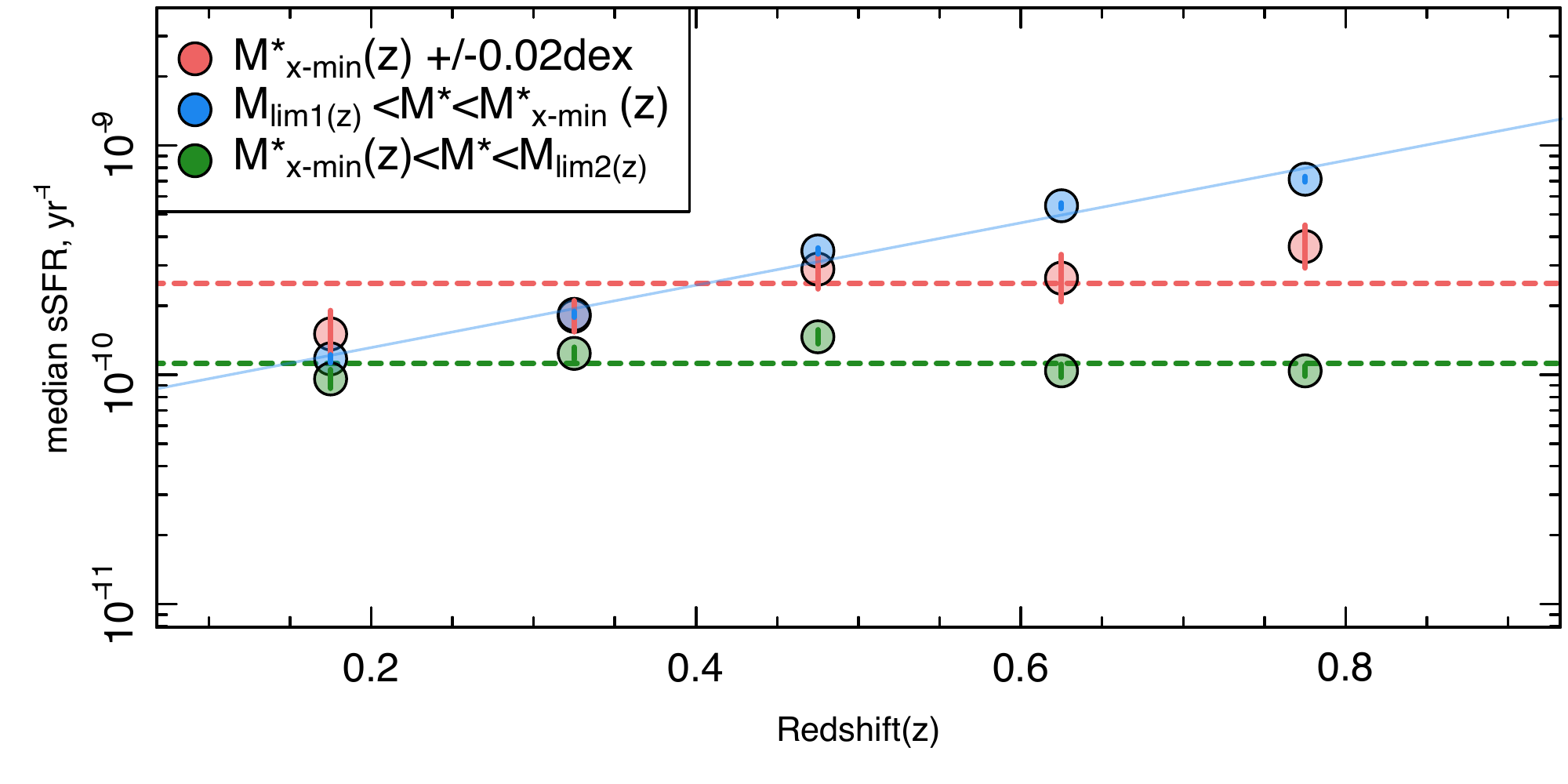}

\caption{Evolution of median sSFRs for three different stellar mass ranges defined with respect to M$^{*}_{x-min}$. Red points display the median sSFR at the predicted evolution of M$^{*}_{x-min}$ from Equation \ref{eq:minevol}, blue and green points show the median sSFR below and above M$^{*}_{x-min}$, respectively. We find that sSFRs for the low stellar mass end evolve strongly, while sSFRs at the M$^{*}_{x-min}$ point and high stellar mass end evolve weakly or not at all. This suggests that M$^{*}_{x-min}$ may occur at a fixed sSFR at all epochs (sSFR$_{x-min}\sim$10$^{-9.6}$yr$^{-1}$ - red dashed line).}
\label{fig:sSFREvol}
\end{center}
\end{figure*}

However, following this train of thought, it is useful to explore how the energy input into the system via star-formation per unit stellar mass ($i.e.$ to first order $\sim$sSFR) evolves in different regimes along the SFS, now defined \textit{relative to} the M$^{*}_{x-min}$ point. If the above postulations are correct, then we may hypothesise that: \\
\\

\noindent  i) The median sSFR of the SFS will evolve strongly with redshift, decreasing as the Universe grows older. This is well known and observed in many previous studies, showing the overall decline of star-formation in the Universe as available gas supply becomes more limited and downsizing occurs. This forms our reference point, with which to compare to other positions within the plane. \\    

\noindent  ii) The median sSFR at M$^{*}_{x-min}$ point will evolve very little with redshift. We are arguing here that this point represents a minimum stellar mass where the energy input into the system from star-formation per unit stellar mass is low enough such that the galaxy can remain in a self-regulated stable state on the SFS without stochastic processes driving it either above (star-bursts) or below (quenching events) the sequence - leading to increased dispersion. As this point may be purely defined by the current SFR and stellar mass of the galaxy (energy input vs gravity to retain gas), one might expect the sSFR at this point to be fixed with time. $i.e.$ as a galaxy grows in stellar mass, it requires more energy input from star-formation to induce stochastic processes that can significantly move a galaxy in the SFR-M$_{\star}$ plane (increasing dispersion). If both stellar mass and this required energy input increase in a self-similar fashion, then we may expect sSFRs at M$^{*}_{x-min}$ to remain fixed. This is a very board assumption as stellar mass $\ne$ gravitational potential (particularly for low mass dwarf galaxies). However, without information regarding each galaxy's dark matter halo, we use it as a proxy here.\\

\noindent  iii) The median sSFR at above and below M$^{*}_{x-min}$ point to evolve differently with redshift. As we propose this point the be the transition between stochastic stellar processes causing large dispersion along the SFS at the low stellar mass end, and large-scale quenching events and the passive population causing large dispersion at the high mass end, we may expect these regimes to evolve differently. Different astrophysical processes are occurring in this regimes, and therefore we may expect sSFRs to also evolve differently. Below M$^{*}_{x-min}$, galaxies are evolving rapidly, gas reservoirs are consumed quickly, and SFRs can change on short timescales. Above M$^{*}_{x-min}$ galaxies are likely globally unaffected by individual stellar events, they evolve slowly unless a catastrophic event ($i.e.$ AGN) causes them to quench and drop off the SFS. This essentially reiterates what is seen in Figure \ref{fig:SFSevol}. However, care must be taken as environment will likely play a strong role in governing these processes, and we reiterate that the environmental impact on the results presented here will be the subject of a following paper.\\

In Figure \ref{fig:sSFREvol} we test these hypotheses by showing the evolution of median sSFR for samples selected with respect to M$^{*}_{x-min}$. First we take the functional form of the predicted evolution of M$^{*}_{x-min}$ from Equation \ref{eq:minevol} and determine the median sSFR at M$^{*}_{x-min}\pm0.02$\,dex at each epoch, shown as the red points. Interestingly, as predicted the sSFR at M$^{*}_{x-min}$ shows no, or very weak evolution with redshift. This suggests that M$^{*}_{x-min}$ is essentially tracing a fixed point in sSFR at all times, $i.e.$ the minimum dispersion along the SFS, while occurring at different stellar masses, always occurs at the same sSFR at close to sSFR$\sim$10$^{-9.6}$yr$^{-1}$ (red dashed line). We therefore argue that galaxies with a sSFR$\sim$10$^{-9.6}$yr$^{-1}$, which we now define as sSFR$_{x-min}$, at all epochs reside in the most stable state on the tight SFS. We reiterate that they are massive enough to retain a steady supply of gas for future star-formation and do not have SFRs large enough to impact their position relative to the SFS through starburst event, while the are not yet massive enough to have formed a powerful AGN which can lead to galaxy-wide quenching. At lower stellar masses, stochastic stellar processes cause galaxies to evolve rapidly producing larger SFR dispersions and the most rapid evolution in sSFRs (blue points/lines in Figure \ref{fig:sSFREvol}). At higher stellar masses, galaxies are in a relatively stable state until catastrophic quenching events drive them off the SFS. As such, median sSFRs (and SFR dispersions) are also relatively flat with time. Figure \ref{fig:sSFREvol} therefore also shows that the SFS above M$^{*}_{x-min}$ is evolving more slowly than below M$^{*}_{x-min}$ ($i.e.$ the slope of the SFS is changing). This is consistent with the results outlined in \cite{Thorne21} who find show that the slope of the low stellar mass end of the SFS is evolving more rapidly than the high stellar mass end and the results presented in Figure \ref{fig:SFSevol}.


Finally, based on our hypothesis and results, we suggest that sSFR$_{x-min}$ represents a fundamental property of galaxies at all epochs. It is the maximum sSFR at which a galaxy can remain in a steady state on the tight star-forming main sequence. If sSFRs are larger than sSFR$_{x-min}$, stochastic stellar processes can push galaxies both above and below the tight sequence leading to increased SFR dispersion. As the Universe evolves the sSFR of all galaxies decreases. As such, another intriguing potential consequence of fixed sSFR$_{x-min}$ point is that with time, a larger fraction of galaxies will drop below the sSFR$_{x-min}$ point. This not only causes M$^{*}_{x-min}$ to move to lower stellar masses (due to the slope of the sSFR-M$_{\star}$ relation) but also predicts that the low stellar mass end of the SFS should become tighter with time as more and more galaxies at lower and lower stellar masses are not strongly impacted by stochastic stellar processes. This is somewhat observed in our Figures \ref{fig:Msig}, \ref{fig:scatevol} and \ref{fig:dists}, as the dispersion at low stellar masses decreases with time - adding weight to this idea. Following this, it is also likely that the $\sigma_{\mathrm{SFR}}$-M$_{\star}$ relation should become more broadly `U-shaped' and less `V-shaped' with time, as more of the low scatter plateau of self-regulated star-formation is revealed when M$^{*}_{x-min}$ moves to lower and lower stellar masses. This is not currently observed in our DEVILS data. However, it is worth noting that the $\sigma_{\mathrm{SFR}}$-M$_{\star}$ relation observed for GAMA at $z\sim0$ from \cite{Davies19a} does have a much broader low-dispersion plateau than our results at higher redshifts. Despite this, further samples extending to lower stellar masses in the local Universe will be required to fully map the local $\sigma_{\mathrm{SFR}}$-M$_{\star}$ relation and its late-time evolution $i.e.$ from 4MOST-WAVES-wide \citep{Driver19}. Further, this potentially means that at some point in our Universal future, as sSFRs continue to fall, all galaxies will sit below the sSFR$_{x-min}$ point. At this time all star-forming galaxies will lie on a very tight, low-dispersion SFS until their gas supply is expended and they asymmetrically fall into the passive regime, $i.e.$ our Universal history is tending to a state where stochastic stellar processes are becoming less and less important in terms of the overall evolution of galaxies.

In combination, our results and speculative interpretations all suggest that the shape of the $\sigma_{\mathrm{SFR}}$-M$_{\star}$ relation, and importantly the M$^{*}_{x-min}$ and sSFR$_{x-min}$ points, are incredibly important diagnostics of the galaxy population and encode detailed information regarding the astrophysical processes with drive the position of galaxies in the SFR-M$_{\star}$ plane. As such, these metrics require further study with larger/deeper samples, and an exploration of how they vary with other galaxy properties such as AGN fraction, morphology/structure, larger-scale environment and merger events. Following this, it is also interesting to consider why galaxies reside in high-dispersion region of this plane by exploring their short- and long-duration star-formation history \citep[$i.e.$][, Throne et al in prep]{Bellstedt20} and/or likely evolutionary path. Each of these will be the subject of further papers using the DEVILS sample.

\section{Conclusions}

In this work we have explored the SFR dispersion as a function of stellar mass across the SFR-M$_{\star}$ plane ($\sigma_{SFR}$-M$_{\star}$) at a number of epochs using the \textsc{ProSpect} SED fitting analysis to DEVILS galaxies. We find that dispersion follows a characteristic `U-shape' at all epochs, with high SFR dispersion at both low and high stellar masses and a minimum SFR dispersion point between (Figure \ref{fig:Msig}). We define this minimum SFR dispersion point as M$^{*}_{x-min}$ and show that it evolves with redshift, moving to higher stellar masses with increasing lookback time (Figure \ref{fig:redevol}), and the absolute SFR dispersion value at M$^{*}_{x-min}$  increases to earlier times (Figure \ref{fig:scatevol}). We also show that the shape of the $\sigma_{SFR}$-M$_{\star}$ relation evolves with time with the SFR dispersion decreasing rapidly at low stellar masses (galaxies are reducing in stochasticity or the diversity in SFHs is decreasing) but increasing for high stellar mass galaxies (more massive galaxies are becoming passive with time), shown in Figure \ref{fig:scatevol}.     

We then apply both colour and sSFR cuts to our sample to compare to existing observations and simulation (which apply similar cuts), and find roughly broad agreement with previous work (Figure \ref{fig:scatevolComp}) - with the caveat of these results being largely dependant on the methodology used. Following this we explore the full distribution of sSFRs as a function of stellar mass, defined relative to M$^{*}_{x-min}$ (Figure \ref{fig:dists}) and find that the distribution of sSFRs evolves different above and below M$^{*}_{x-min}$, with strong evolution at the low stellar mass end, and little evolution above M$^{*}_{x-min}$. Exploring this further we showed that M$^{*}_{x-min}$ does appear to trace the turnover point in the star-forming sequence (Figure \ref{fig:SFSevol}), potentially indicating that it traces the boundary between two different evolutionary regimes.  

We place our new observational results in the context of existing astrophysical interpretations of the origin of the shape of the $\sigma_{SFR}$-M$_{\star}$ relation, that the low mass scatter is caused by stochastic SF process, while the high mass scatter is caused by AGN feedback, and show that they are at least consistent with this interpretation. However, we caveat that they do not rule out many other interpretations such at long-duration SFH variation, environmental impacts, morphology/structure variations across the plane, etc. These will all be discussed in future papers using the DEVILS sample, and hence we do not discuss them further here. 

We finally highlight that, as the M$^{*}_{x-min}$ point occurs at roughly a fixed sSFR all epochs and traces the turnover point in the SFS, it is likely a key parameter in our understanding of galaxy evolution processes, and potentially delineates the boundary between different evolutionary mechanism (whatever the ultimately turn out to be). As such, and investigation into the relationship between the evolution of M$^{*}_{x-min}$ and other galaxy properties will likely yield signifiant insights into the galaxy evolution process. These investigations will form the basis of a series of papers exploring M$^{*}_{x-min}$ within the DEVILS sample.

\section*{Acknowledgements}

 LJMD, ASGR and LC  acknowledge support from the \textit{Australian Research Councils} Future Fellowship scheme (FT200100055, FT200100375 and FT180100066, respectively). JET is supported by the Australian Government Research Training Program (RTP) Scholarship. SB and SPD acknowledge support from the \textit{Australian Research Councils} Discovery Project scheme (DP180103740). MS has been supported by the European Union's Horizon 2020 Research and Innovation programme under the Maria Sklodowska-Curie grant agreement (No. 754510), the Polish National Science Centre (UMO-2016/23/N/ST9/02963), and the Spanish Ministry of Science and Innovation through the Juan de la Cierva-formacion programme (FJC2018-038792-I). MJJ acknowledges  support from the UK Science and Technology Facilities Council [ST/S000488/1 and ST/N000919/1], the Oxford Hintze Centre for Astrophysical Surveys which is funded through generous support from the Hintze Family Charitable Foundation and the South African Radio Astronomy Observatory (SARAO), and ICRAR for financial support during a sabbatical visit. MB acknowledges the support of the University of Western Australia through a Scholarship for International Research Fees and Ad Hoc Postgraduate Scholarship. Parts of this research were conducted by the Australian Research Council Centre of Excellence for All Sky Astrophysics in 3 Dimensions (ASTRO 3D), through project number CE170100013. DEVILS is an Australian project based around a spectroscopic campaign using the Anglo-Australian Telescope. DEVILS is part funded via Discovery Programs by the \textit{Australian Research Council} and the participating institutions. The DEVILS website is https://devilsurvey.org. The DEVILS data is hosted and provided by AAO Data Central (\url{datacentral.org.au}). 

\section{Data Availability}
Data products used in this paper are taken from the internal DEVILS team data release and presented in \cite{Davies21a} and \cite{Thorne21}. These catalogues will be made public as part DEVILS first data release described in Davies et al (in prep).

\appendix

\section{Example SFR Error Distribution}
\begin{figure}
\begin{center}
\includegraphics[scale=0.65]{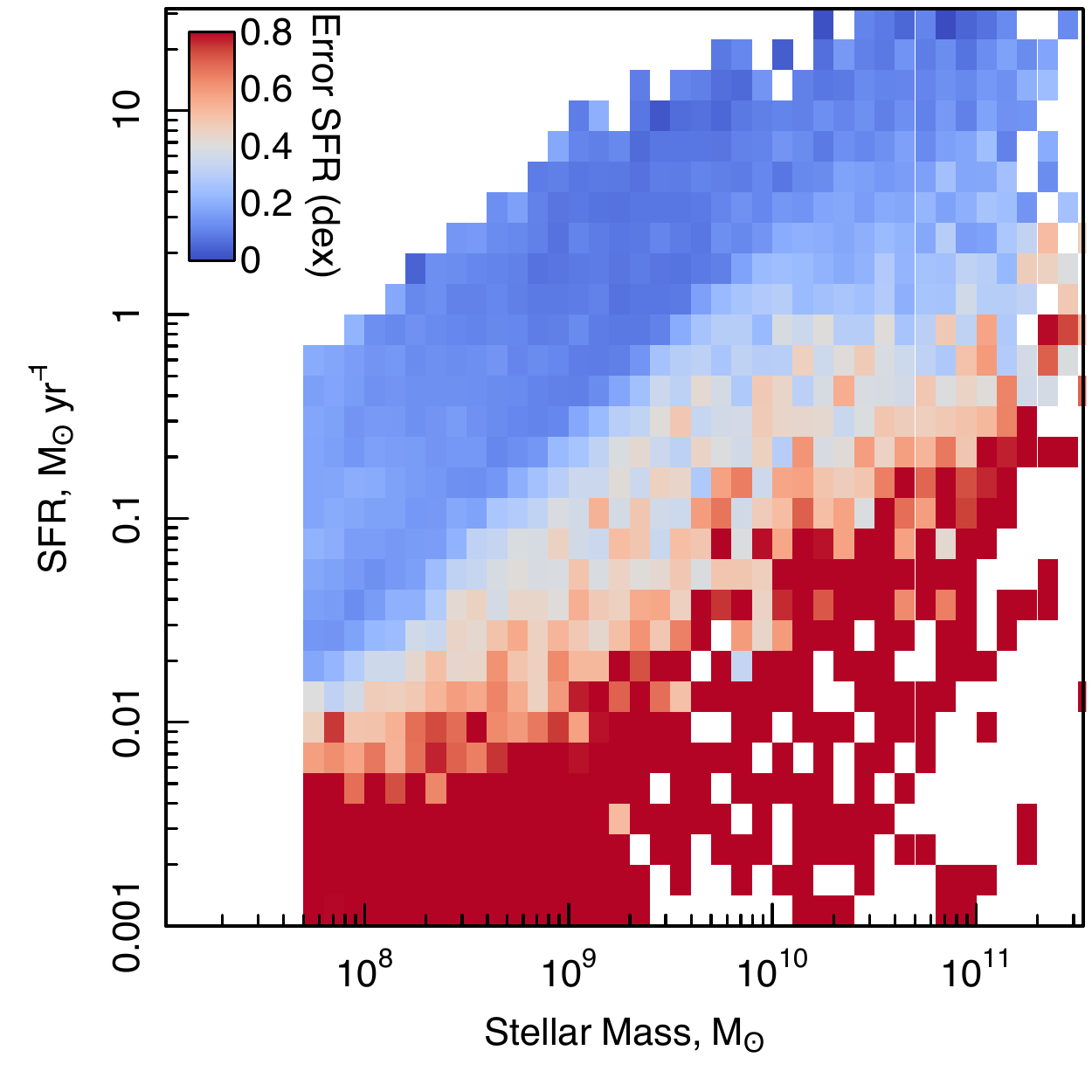}
\caption{The typical median SFR error from the \textsc{ProSpect} analysis of \citet{Thorne21} as a function stellar mass and SFR for our $0.4<z<0.55$ sample. Errors along the SFS are $<0.2$\,dex at all stellar masses. }
\label{fig:errors}
\end{center}
\end{figure}

In order to aid in understanding of the significance of our results, in Figure \ref{fig:errors} we show the median SFR errors from the \textsc{ProSpect} analysis of \cite{Thorne21} as a function of stellar mass and SFR in our $0.4<z<0.55$ sample. We find errors of $<0.2$\,dex for the SFR population at all stellar masses.

\section{Detailed comparison to previous work}   

Firstly, in order to compare our results to previous results exploring the dispersion along the SFS, we measure the dispersion for both our  sSFR cut and colour cut samples described in Section \ref{sec:sigma_M_cut} at the three stellar mass points (log$_{10}$[M$_{\star}$/M$_{\odot}$]=8.5, 9.5, 11.0). These mass points match the EAGLE work of \cite{Katsianis19}, and others. Figure \ref{fig:scatevol} shows the evolution of dispersion of the SFS in these three stellar mass bins as the filled purple, black and red symbols, respectively. Here in the top panel we show $\sigma_{SD}$ dispersion metric using our NUV-r colour cuts, while in the bottom panel we display the intrinsic dispersion measurements using the sSFR cut. This choice is to best compare to previous observations (top) and simulations (bottom), which largely use similar selection criteria (see following subsections). We then fit a linear relation to each of the points with the solid lines in the corresponding colour. 

In the top panel of Figure \ref{fig:scatevol}, for $\sigma_{SD}$ and colour cuts, we find strong evolution in the low (log$_{10}$[M*/M$_{\odot}$]=8.5) and intermediate (log$_{10}$[M*/M$_{\odot}$]=9.5) stellar mass bins, with the dispersion increasing with redshift. While in our highest stellar mass bin (log$_{10}$[M*/M$_{\odot}$]=11) we find that the dispersion is largely constant at around 0.4-0.5\,dex. In the bottom panel for intrinsic scatter and sSFR cuts, we find a declining or constant dispersion at all stellar masses, ranging from 0.45\,dex to 0.15\,dex.  However, we must also caveat again that these relations are largely dependant on the choice of selection used to isolate the SFS population, and are purely used here to compare to previous results that apply similar selections.

\subsection{Comparison to previous observations}    
\label{sec:prev}    

Numerous observational studies have measured the dispersion of the SFS at various epochs and over a range of different stellar mass scales. Here we compile these results and match to the most appropriate stellar mass point in Figure \ref{fig:scatevol} for comparison. However, it is worth noting that the majority of these studies only probe the high mass end of the $\sigma_{\mathrm{SFR}}$-M$_{\star}$ relation, above the M$^{*}_{x-min}$  point. This means that they are only directly comparable to our log$_{10}$[M$_{\star}$/M$_{\odot}$]=11 measurements and are also in the region of the SFR-M$_{\star}$ plane where the measurement of the dispersion is most sensitive to the choice of passive/star-forming selection. That said, it is still informative to compare our new measurements to these previous results.    

Firstly, \cite{Noeske07} use the All-Wavelength Extended Groth Strip International Survey (AEGIS) with Keck/DEEP2 spectra to measure the SFS from $0.2<z<1.1$ at log$_{10}$[M$_{\star}$/M$_{\odot}$]$>$10. They measure the SFR dispersion about the main-sequence to be a constant $\sim0.3$\,dex at all epochs, assuming a $log-normal$ distribution. They determine SFRs using a combination of the DEEP2 emission lines, GALEX UV photometry and Spitzer 24\,$\mu$m emission, and select star-forming galaxies based on U-B colours. Stellar masses are derived from SED fits. In Figure \ref{fig:scatevol} we display the \cite{Noeske07} values as red crosses as they are most directly comparable to the high stellar mass end of our sample. These points lie close (with small normalisation offset) to our measured values, which at log$_{10}$[M$_{\star}$/M$_{\odot}$]$=$11 and for our colour selected sample show a flat $\sim$0.4\,dex dispersion at $0<z<0.8$ 

Next, \cite{Guo13} use the COSMOS data and publicly available photometric redshifts to measure the dispersion of the SFS at $0.6<z<0.8$. They also determine sSFRs using the UV+24\,$\mu$m measurements from the COSMOS photometric sample, and stellar masses using a simple scaling from rest-frame K-band emission. Star-forming galaxies are selected using a U, V, K colour magnitude diagram. They measure the dispersion at a single epoch but within four $\Delta$stellar mass bins between 9.5$<$log$_{10}$[M*/M$_{\odot}$]$<$11.5, finding a sSFR dispersion of 0.18, 0.21, 0.26, 0.31\,dex respectively. In Figure \ref{fig:scatevol} we display the \cite{Guo13} results for the 9.5$<$log$_{10}$[M*/M$_{\odot}$]$<$10.0 and 11.0$<$log$_{10}$[M*/M$_{\odot}$]$<$11.5 bins in comparison to our intermediate and high stellar mass points, the open black and red circles at $z=0.7$. The high stellar mass scatter point is consistent with the results of \cite{Noeske07}, but at a lower normalisation than our point at that redshift. However, the 9.5$<$log$_{10}$[M*/M$_{\odot}$]$<$10.0 point lies well below our current measurement. Importantly, at that point in the \cite{Guo13} sample, the sample is dominated by objects not detected at 24\,$\mu$m and as such, the sample may be incomplete at the low SFR end - leading to a decrease in measured dispersion.

Following this, \cite{Guo15} further measured the dispersion along the main-sequence for $z\sim0$ galaxies using a sample derived from SDSS. Here they measure SFRs using a combination of H-$\alpha$ emission and $WISE$ 24\,$\mu$m photometry. The star-forming population is selected using a sSFR cut and dispersion measured for both the full population and disk galaxies only (using a number of different morphological selections).  They measure the dispersion to have a minimum point close to log$_{10}$[M$_{\star}$/M$_{\odot}$]$=$9.0, comparable to the results from \cite{Davies19b}, and a minimum scatter of 0.4\,dex and 0.51\,dex at log$_{10}$[M$_{\star}$/M$_{\odot}$]$=$9.5 and 11.5, respectively. In Figure \ref{fig:scatevol} we display these results as the open upward-facing triangles close to $z=0$, and find that they are largely consistent with our evolutionary trends (and roughtly consistent with the GAMA $z\sim0$ points).

\cite{Ilbert15} also explore the evolution of the mass-sSFR plane from $z=1.4$ to the present day using the COSMOS/GOODS samples. They primarily select galaxies based on 24\,$\mu$m emission, and determine SFRs using MIR+FIR photometry. Stellar masses are estimated using the $Le$ $Phare$ SED fitting code and star-forming galaxies selected using rest-frame colour. They measure $\sigma_{\mathrm{MS}}$ at a number of epochs and for a number of stellar mass bins at log$_{10}$[M$_{\star}$/M$_{\odot}$]>10. Here we use their 10.5<log$_{10}$[M*/M$_{\odot}$]$<$11.0 stellar mass bin to compare to our results, which is shown as the red crosses in  Figure \ref{fig:scatevol}. These display a measured dispersion that is comparable to ours and no evolution with redshift, also consistent with our findings at this stellar mass range.

\cite{Schreiber15} use the GOODS-$Herschel$ and CANDELS-$Herschel$ key programs to explore the growth of galaxies from $z=4$ to the present day. The derive SFRs from the UV+FIR and stellar masses from SED fitting. Star-forming galaxies are selected using rest-frame colours. They measure the SFS dispersion at a number of epochs and for a number of stellar mass bins at log$_{10}$[M$_{\star}$/M$_{\odot}$]$>$9.5. In Figure \ref{fig:scatevol} we display their log$_{10}$[M$_{\star}$/M$_{\odot}$]=9.5 and log$_{10}$[M$_{\star}$/M$_{\odot}$]=11 measurements at $z=0.5$ (the only ones that overlap with our sample) as the open back and red diamonds respectively. \cite{Schreiber15} find a largely consistent dispersion value at all stellar masses, which is roughly consistent with our measurements at the same epoch - 0.35\,dex compared to $\sim$0.4\,dex.

More locally, \cite{Willett15} use the Galaxy Zoo sample to measure the $\sigma_{\mathrm{SFR}}$-M$_{\star}$ relation at $z<0.085$ for different morphologically-selected disk populations from 8.5$<$log$_{10}$[M*/M$_{\odot}$]$<$11.5. They use SFRs and stellar masses from the SDSS sample, and select disk galaxies based on the Galaxy Zoo classifications. They also find the `U-shape' distribution of the $\sigma_{\mathrm{SFR}}$-M$_{\star}$ relation. Here we use their values at each of our mass bins and display them in Figure \ref{fig:scatevol} as open downward facing triangles at $z\sim0$.  At all masses these points are largely consistent with our trends. However, they do find that the smallest scatter is in the log$_{10}$[M$_{\star}$/M$_{\odot}$]=11 bin (the up-turn in dispersion in their sample occurs at higher stellar masses than ours).       

Finally \cite{Boogaard18} use the MUSE Hubble Ultra Deep Field Survey to measure the evolution of the SFR-M$_{\star}$ relation to $z=1$. Stellar masses are derived from SED fitting and SFR from MUSE spectral lines. They measure the SFR dispersion to be constant at 0.44 for all redshifts and stellar masses. As such, we simply display this in Figure \ref{fig:scatevol} as the horizontal dashed gold line. This value sits between all of our stellar mass bins and is therefore consistent with the median at all stellar masses.     

From this comparison we can see that the observation picture is complicated and varied. However, despite the samples being selected in very different ways and the $\sigma_{\mathrm{SFR}}$-M$_{\star}$ relation measured with varied approaches, our new measurements are largely consistent with the existing literature. The only exception to this is the log$_{10}$[M$_{\star}$/M$_{\odot}$]$=$9.5 point from \cite{Guo13} at $z=0.5$, which is in strong contention with our new measurements. However, we note again that this measurement is made at a point where the majority of their sample is undetected in the observational band used to measure their SFRs.   

\subsection{Comparison to EAGLE Hydrodynamical Simulations}    
\label{sec:psims}

Recently, various simulation suites have also been used to estimate the evolution of the $\sigma_{\mathrm{SFR}}$-M$_{\star}$ relation. Firstly, \cite{Matthee18} use Evolution and Assembly of GaLaxies and their Environments \cite[EAGLE,][]{Schaye10,Crain15} hydrodynamical simulations to measure the evolution of $\sigma_{\mathrm{MS}}$ for sSFR$>$10$^{-11}$\,yr$^{-1}$ galaxies at 9$<$log$_{10}$[M*/M$_{\odot}$]$<$11 and find a dispersion that deceases roughly linearly with stellar mass (as in our sample for a similar sSFR cut, see Figure \ref{fig:Msig_cut}), but that declines relatively uniformly at all stellar masses out to $z=1$. In the bottom panel of Figure \ref{fig:scatevol} we display the \cite{Matthee18} evolution of the dispersion along the SFS (taken from their Figure 3) in each of our stellar mass bins as the short dashed lines. We note that these lines display the $intrinsic$ dispersion measurements not including measurement error, and hence we compare to our intrinsic values (filled points and linearly fit with the solid lines). Overall we find very similar trends to \cite{Matthee18} in our sample which is selected in a similar manner. We find a decreasing or flat evolution of the dispersion at all stellar masses for our intrinsic dispersion and absolute dispersion values that lie close to the \cite{Matthee18} lines. We also both find the largest dispersion in the low stellar mass bin and smallest dispersion in the high stellar mass bin. For $z=0$ \cite{Matthee18} also display their values including a SDSS-like measurement error, which we show in the bottom panel of Figure \ref{fig:scatevol} as open triangles. These points are also loosely consistent with our observed trends.

Following this, \cite{Katsianis19} also used EAGLE to explore the evolution of  $\sigma_{\mathrm{SFR}}$-M$_{\star}$, using a similar sSFR selection to \cite{Matthee18}, but find somewhat different results. While \cite{Matthee18} determine that $\sigma_{\mathrm{SFR}}$ decreases with stellar mass, \cite{Katsianis19} recover the 'U-shaped' distribution of $\sigma_{\mathrm{SFR}}$-M$_{\star}$ at a range of epochs but with a decreasing normalisation with redshift. Interestingly, their sSFR cut does not remove the large scatter at the high stellar mass end as we see in our sample \citep[and in][]{Matthee18}. Most importantly \cite{Katsianis19} find the smallest dispersion at intermediate stellar masses, close to our M$^{*}_{x-min}$ points. While this is also true for our samples with no cuts applied and with our colour-selections, it is not for a sSFR cut \citep[as applied in][]{Katsianis19}. However, we caution once again here that the choice of sSFR cut applied can have strong impact on the measured dispersion values, especially at the high stellar mass end. The evolution of $\sigma_{\mathrm{SFR}}$ presented by \cite{Katsianis19} is shown in the bottom panel of Figure \ref{fig:scatevol} as the long dashed lines. Despite the statements above the intrinsic scatter values presented by \cite{Katsianis19} and overall trends are largely consistent with our observations, modulo a normalisation offset in the highest stellar mass bin.       

In summary, the \cite{Matthee18} and \cite{Katsianis19} work highlight how the choice of SFS selection and methodology can strongly affect the measurement of the $\sigma_{\mathrm{SFR}}$-M$_{\star}$ relation (they use the same simulations, but get different results), but to first order none of the EAGLE predictions are in strong contention with our observational trends.

\subsection{Comparison to Shark Semi-Analytic Simulations}    
\label{sec:shark}

\begin{figure*}
\begin{center}
\includegraphics[scale=0.6]{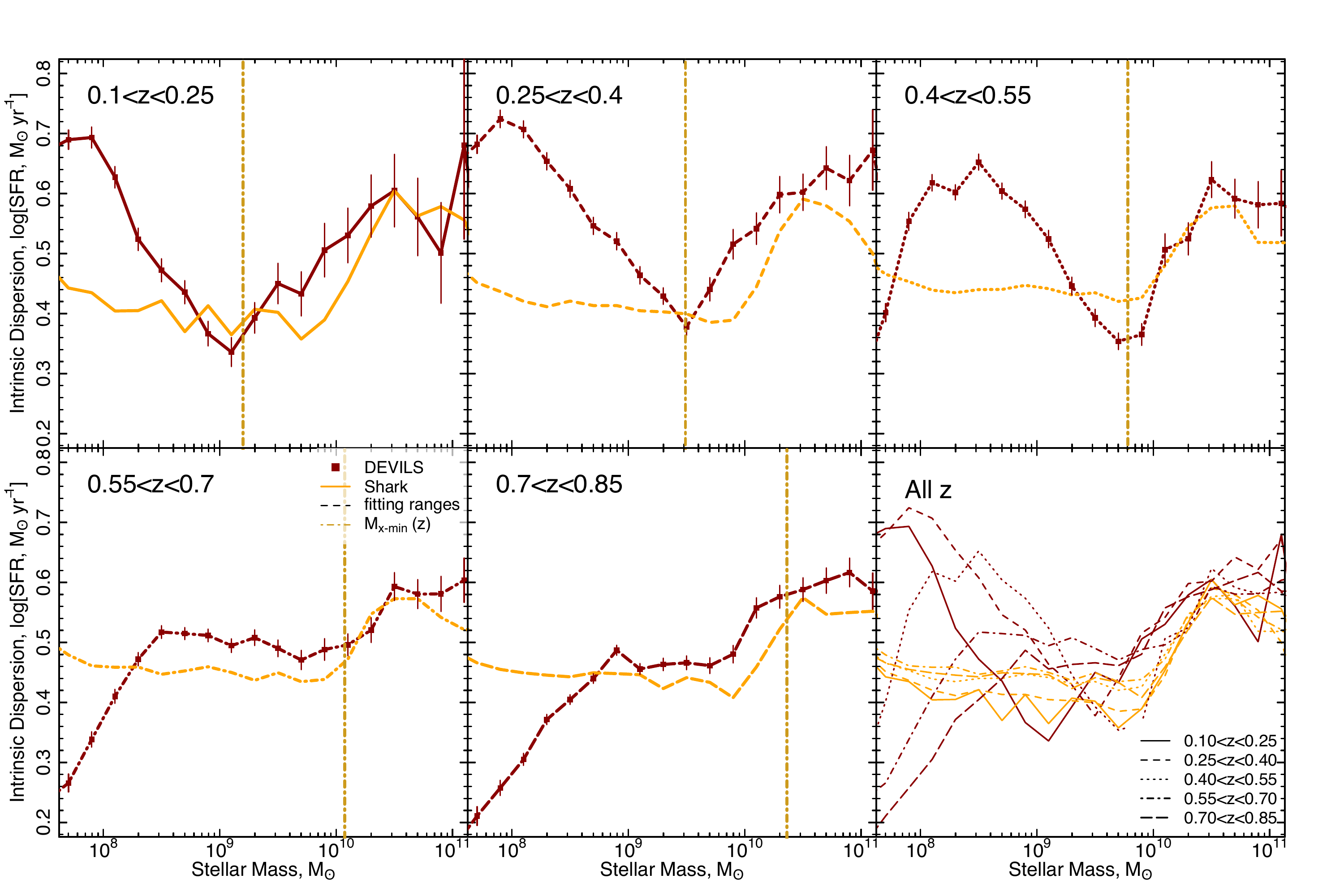}
\caption{Comparison between the intrinsic $\sigma_{\mathrm{SFR}}$-M$_{\star}$ relation and its evolution from our DEVILS observations (dark red) and Shark semi-analytic model (orange), similar to Figure \ref{fig:Msig}. We display this comparison for each of our redshift bins and all redshifts in the same panel (bottom right). We also display the predicted evolution of M$^{*}_{x-min}$ from Equation \ref{eq:minevol} as the dashed vertical gold line.}
\label{fig:SharkCompApp}
\end{center}
\end{figure*}

Lastly, we also compare to the Shark semi-analytic model \citep{Lagos18}. In comparing to Shark we have much more flexibility in directly comparing our observational results to the simulation as within the DEVILS team we have developed bespoke DEVILS-specific light-cones from the Shark simulation suite (Bravo et al, in prep). These have the same sample selection and source distributions as the DEVILS sample and therefore can be directly compared. As such, we apply an identical sample selection and analysis procedure for measuring the $\sigma_{\mathrm{SFR}}$-M$_{\star}$ relation as in our DEVILS sample (as described earlier in this work). We measure the intrinsic scatter in the Shark-simulated population in each of our redshift windows as a function of stellar mass for the full population and identical sSFR selection as our observational data (1\,dex below the Shark SFS cut), and then also determine M$^{*}_{x-min}$, and the intrinsic dispersion at M$^{*}_{x-min}$ and log$_{10}$[M$_{\star}$/M$_{\odot}$]=8.5, 9.5, 11.0. Figure \ref{fig:SharkCompApp} shows an example of the Shark intrinsic $\sigma_{\mathrm{SFR}}$-M$_{\star}$ relation in comparison to our DEVILS measurements.   

We find that while the Shark $\sigma_{\mathrm{SFR}}$-M$_{\star}$ does broadly show the characteristic `U-shape' at all epochs \citep[as reported for $z\sim0$ in][]{Davies19a}, and has similar minimum intrinsic dispersion values as our DEVILS observations (at $\sim0.4$\,dex), we find two interesting differences to our observations. Firstly, the $\sigma_{\mathrm{SFR}}$-M$_{\star}$ relation has a larger flat low dispersion plateaux, and in fact shows consistent $\sim0.4$\,dex dispersion between 8$<$log$_{10}$[M*/M$_{\odot}$]$<$10. Secondly, the Shark relation shows very little evolution with redshift in comparison to DEVILS (see bottom right panel of \ref{fig:SharkCompApp}).  

Next we also include lines from the Shark sSFR-selected SFS dispersion on the bottom panel of Figure \ref{fig:scatevol} as the solid blue lines. These also show very little evolution at any stellar mass and differ to both our observational results and EAGLE, in that even with a sSFR selection the largest SFR dispersion in the Shark sample occurs at the highest stellar masses. This largely occurs as using a sSFR cut defined at 1\,dex below the SFS still retains a significant fraction of passive galaxies in the Shark sample. However, for direct comparison, we do not change the method which we use for the observational data.     

We do note, as above, that the Shark dispersion close to M$^{*}_{x-min}$ ($i.e.$ at log$_{10}$[M*/M$_{\odot}$]=9.5) lies close to our measured observational values when a sSFR selection is applied (black points compared to lowest blue line in the bottom panel of Figure \ref{fig:scatevol}).  We therefore, once again highlight that the Shark SFS minimum dispersion values are close to our observational data, but the evolution of the shape of the $\sigma_{\mathrm{SFR}}$-M$_{\star}$ relation is not consistent.

\bsp	
\label{lastpage}
\end{document}